\documentclass{mn2e}

\usepackage{epsfig,amssymb}

\def\LCDM{{$\Lambda$CDM}}

\def\Mpc{{\,h^{-1}\,{\rm Mpc}}} 
\def\Mpccube{{\,h^{-3}\,{\rm Mpc^3}}} 
\def\lstar{{L$_{\star}$}}
\def\etal{{et~al.}} 
\def\wprp{{w$_{\rm p}$(r$_{\rm p}$)/r$_{\rm p}$}}
\def\wp{{w$_{\rm p}$(r$_{\rm p}$)}}
\def\nsub{{$N_{sub}$}}
\def\nsubm{{N_{sub}}}
\def\nr{{$N_{r}$}}
\def\nrm{{N_{r}}}

\newcommand{\notice}[1]{\textbf{[\textit{#1}]}} 
\newcommand{\plotone}[1] 
           {\centering \leavevmode \psfig{file=#1,width=\columnwidth,clip=}} 
\newcommand{\plottwo}[2] 
           {\centering \leavevmode \psfig{file=#1,width=\columnwidth,clip=} 
                            \hfill \psfig{file=#2,width=\columnwidth,clip=}}

\newcommand{\plotthree}[6]            
    {\centering \epsfig{file=#1,width=#2\textwidth,clip=}
         \hfill \epsfig{file=#3,width=#4\textwidth,clip=}
         \hfill \epsfig{file=#5,width=#6\textwidth,clip=}}

\def\simlt{\lower.5ex\hbox{$\; \buildrel < \over \sim \;$}}
\def\simgt{\lower.5ex\hbox{$\; \buildrel > \over \sim \;$}}

\title[SAGS-I: Robust error estimation for 2-point clustering statistics]
{Statistical Analysis of Galaxy Surveys-I. 
Robust error estimation for 2-point clustering statistics}

\author[P. Norberg, C.M. Baugh, E. Gazta\~naga, \& D.J. Croton]{ 
P. Norberg$^1$, 
C. M. Baugh$^2$, 
E. Gazta\~{n}aga$^3$, 
D. J. Croton$^{4,5}$\\ 
$^1$SUPA\thanks{The Scottish Universities Physics Alliance}, Institute for Astronomy, University of Edinburgh, Royal Observatory, Blackford Hill, Edinburgh, EH9 3HJ, UK.\\
$^2$Institute for Computational Cosmology, Department of Physics, University of Durham, South Road, Durham DH1 3LE, UK.\\
$^3$Instituto de Ciencias del Espacio (IEEC/CSIC), F. de Ciencias UAB, Torre C5- Par-2a, Bellaterra, 08193 Barcelona, Spain.\\
$^4$Department of Astronomy, University of California, Berkeley, CA, 94720, USA. \\
$^5$Centre for Astrophysics and Supercomputing, Swinburne University of Technology, Mail H39, PO Box 218, Hawthorn, Victoria, \\
3122, Australia\\
} 
 
\date{Accepted ---. Received ---;in original form ---} 
 
\begin{document} 
 
\maketitle 
 

\begin{abstract} 
We present a test of different error estimators for 2-point clustering
statistics, appropriate for present and future large galaxy redshift
surveys. Using an ensemble of very large dark matter \LCDM\ N-body 
simulations, we compare internal error estimators (jackknife
and bootstrap) to external ones (Monte-Carlo realizations). For
3-dimensional clustering statistics, we find that none of the internal
error methods investigated are able to reproduce neither accurately 
nor robustly the errors of external estimators on 1 to $25\Mpc$ scales. 
The standard bootstrap overestimates the variance of $\xi(s)$ by 
$\sim 40$\% on all scales probed, but recovers, in a robust fashion, 
the principal eigenvectors of the underlying covariance matrix. The 
jackknife returns the correct variance on large scales,
but significantly overestimates it on smaller scales. This scale
dependence in the jackknife affects the recovered eigenvectors, which 
tend to disagree on small scales with the external estimates.  
Our results have important implications for the use of galaxy 
clustering in placing constraints on cosmological parameters. 
For example, in a 2-parameter fit to the projected correlation 
function, we find that the standard bootstrap systematically 
overestimates the 95~\% confidence interval, while the jackknife 
method remains biased, but to a lesser extent. The scatter we 
find between realizations, for Gaussian statistics, 
implies that a 2-$\sigma$ confidence interval, as inferred from an 
internal estimator, could correspond in practice to anything from 
1-$\sigma$ to 3-$\sigma$.  
Finally, by an oversampling of sub-volumes, it is possible to obtain 
bootstrap variances and confidence intervals that agree with external 
error estimates, but it is not clear if this prescription will work 
for a general case.
\end{abstract} 
                                   

\section{Introduction}
\label{sec:intro}

The large-scale structure (LSS) of the Universe, as traced by galaxies  
and
clusters, encodes important information about the basic cosmological
parameters and also the physical process which underpin the formation of
cosmic structures. LSS measurements are now routinely used in  
conjunction
with measurements of the cosmic microwave background (CMB) to constrain
cosmological parameters (e.g. Cole \etal\ 2005; Sanchez \etal\ 2006;  
Tegmark
\etal\ 2006; Percival \etal\ 2007). However, the estimation of errors  
in CMB
measurements is, by comparison with LSS work, quite sophisticated and
rigorous.

To constrain the cosmological world model from the galaxy distribution, 
large volumes and  galaxy
numbers are needed. The current generation of local galaxy surveys,  
such as
the 2dFGRS (Colless \etal\ 2001) and SDSS (York \etal\ 2000) satisfy 
both of these requirements 
and can be divided into large sub-samples to study clustering  
trends
with galaxy properties (e.g. Norberg \etal\ 2001, 2002; 
Zehavi \etal\  2002, 2004, 2005; Madgwick \etal 2003;
Croton \etal\ 2004, 2007; Gazta\~naga \etal\ 2005; Li \etal\
2006, 2007). In anticipation of even bigger surveys at intermediate  
and high
redshifts (e.g. GAMA, VVDS-CFHTLS, Euclid), it is timely to revisit the
techniques used to estimate errors on clustering statistics from galaxy
redshift surveys. In particular, we explore ways to robustly quantify
reliable errors estimates on two-point clustering statistics in 
3-dimensional space. This has important consequences for the
determination of the 
cosmological parameters, both in terms of setting values and in the  
number of
free parameters needed to describe the data, and in uncovering trends  
which
can be used to constrain galaxy formation models.

In a perfect world, the optimal way to estimate the error on a  
clustering
measurement would be to generate a large sample of independent mock  
galaxy
catalogues that look the same as the data (at least in so far as the  
observed
data may be affected by sampling fluctuations, e.g. Efstathiou \& Moody
2001). The challenge here is to ensure that the mocks are a faithful
reproduction of the observations. This is a more demanding requirement  
than
one might at first think. In order to estimate accurate errors on the
two-point statistics, it is necessary that the mocks also reproduce the
higher order clustering displayed by the data, since the errors on  
2-point
statistics implicitly depend on these higher order moments. This  
approach
will inevitably involve N-body simulations, in order to accurately  
model the
underlying dark matter, and so becomes computationally expensive.  
Various
techniques have been adopted in the literature to populate such  
simulations
with galaxies (for a survey see Baugh 2008). The number of simulations
required is large, running into several tens to get reasonable  
estimates of
the variance and hundreds or even thousands to get accurate covariance
matrices.

An alternative empirical method is to use the observed data itself to  
make an
estimate of the error on the measurement. Such ``internal'' error  
estimates
use a prescription to perturb the data set in some way in order to make
copies. These copies allow the statistical distribution which  
underlies the
data to be probed without having to assume anything about its form. In  
this
paper we investigate the performance of two common approaches, the  
jackknife
and bootstrap internal error estimates; both will be compared to the  
errors
calculated from a large suite of simulated mock data sets.

The jackknife method was developed as a quite generic statistical tool  
in the
1950s (Quenouille 1956; Tukey 1958). The bootstrap method (e.g. Efron  
1979)
is a modification or extension of the jackknife made possible by the
availability of fast computers in the 1970s. These internal estimators  
first
divide the data set into subsamples, which consist either of individual
objects or groups of objects, which are then resampled in a particular  
way
(see Section 2 for a full description of the procedure). The first
applications of either technique to astronomical observations date  
from the
early 1980s with the analysis of the velocities of galaxies in clusters
(Lucey 1979; Bothun \etal\ 1983). The bootstrap method was first  
applied to
galaxy clustering by Barrow, Bhavsar \& Sonoda (1984; see also Ling,  
Frenk \&
Barrow 1986).

In these early works sub-samples were simply taken using the individual
objects themselves. However, as pointed out by Fisher \etal\ (1994),  
this can
lead to unreliable errors. From a Monte-Carlo analysis using a series of
N-body simulations, Fisher \etal\ showed that when sampling individual
galaxies the bootstrap method underestimates the error on the density
estimate in voids and, likewise, overestimates it in clusters. Mo,  
Jing \&
B\"orner (1992) present an analytic estimate for the errors on 2-point
statistics and show that the bootstrap errors obtained by resampling  
using
individual galaxies give incorrect errors compared with the ensemble  
average
under certain conditions. These valid objections can be avoided by  
resampling
the data set divided into sub-volumes instead of individual galaxies.  
Hence,
in this paper, we generate copies of data sets by selecting sub-volumes of
the data set.

Resampling of patches or regions of surveys has already been applied  
to error
estimation in LSS analyses. The jackknife technique has been  
extensively used
in projection, including the angular clustering of galaxies (Scranton  
\etal\
2002), in the analysis of CMB maps (e.g. Gazta\~naga \etal\ 2003) and,
perhaps most heavily, in efforts to detect the Integrated Sachs-Wolfe  
effect
by cross-correlating sky maps of galaxy or cluster positions with  
temperature
fluctuations in the cosmic microwave background (Fosalba, Gazta\~naga \&
Castander 2003; Fosalba \& Gazta\~naga 2004; Cabr\'e \etal\ 2007). In
projection, the jackknife variance agrees well with Monte-Carlo  
estimates
from mock catalogues (e.g. Cabr\'e \etal\ 2007). This is perhaps an  
easier
case than the three dimensional surveys dealt with in this paper,  
since the
distributions of galaxy fluctuations will tend to look more Gaussian in
projection (although distinct non-Gaussian behaviour is clearly  
evident in
the projected galaxy fluctuations, e.g. Gazta\~naga 1994).

The jackknife method has also been applied by many authors to the  
estimation
of errors on galaxy clustering in three dimensions using volume  
resampling.
Zehavi \etal\ (2002) carried out a simple test of the accuracy of the
jackknife estimate of the variance and found that the jackknife  
produced an
essentially unbiased estimate, but with a scatter that could approach  
50\%.
Here we carry out a more exhaustive comparison  of
internal error estimators with the ``external'' Monte-Carlo simulation
methods.

The paper is arranged in the following way. Section~\ref{sec:err_est}  
defines
the three error estimators we use throughout, while Section~ 
\ref{sec:err_met}
introduces the simulations which will be used to generate fake data  
sets for
our numerical experiments, along with the clustering statistics and the
principal component decomposition, needed for the error estimator  
comparison.
Section~\ref{sec:err_anal} presents the raw results from three different
error techniques and remains rather technical. We consider in
Section~\ref{sec:test_case} a simple test case scenario, where the
implications of using different error estimators for a  
``straightforward''
two parameter fit to the projected correlation function are inferred. We
summarize our findings in Section~\ref{sec:conclusion}.

\section{Review of error estimation methods}
\label{sec:err_est}

There are numerous ways to estimate the error on a clustering
measurement. In this section, we give an outline of four of the most
popular non-parametric methods in use in the literature, three of which we
compare in this paper. We do not consider in this paper analytic error
estimates (like Poisson), nor parametric error estimates, like those
derived from Gaussian or Log-Normal density fields. The latter
methods are commonly used for estimating errors in the linear
clustering regime (see e.g. Percival \etal\ 2001 and Cole \etal\ 2005
for applications to power spectrum estimates).
Below, we describe three ``internal'' methods  
which use the data itself to derive an estimate of the error on 
a measurement, the sub-sample, jackknife and bootstrap techniques 
(sections~\ref{sec:data_errsubsample}, \ref{sec:data_errJK}
and~\ref{sec:data_errBS}). 
Then we describe in section~\ref{sec:mock_err} the commonly used
``external'' method of creating Monte-Carlo realizations or
reproductions of the data. We close by giving an overview of the
advantages and disadvantages of each method in
Section~\ref{sec:overview}. 

Each of the techniques involves performing measurements on 
copies of the data in order to sample the underlying probability 
distribution of the quantity we are trying to measure. The internal 
estimates make copies or resamplings from the observed data whereas 
the external method generates fake data sets, without manipulating the
observed data. Three assumptions are implicit in the internal 
or ``data inferred'' approaches: 
1) the data gives an accurate representation of the underlying 
probability distribution of measurements; 2) the sub-samples into 
which the data is split are sufficient in number to allow accurate 
enough estimates of the errors and 3) the sub-sample volumes are large 
enough to be representative. Condition 1 would be violated if a second 
independent measurement from an equivalent data set gave a significantly 
different estimate of the measurement. In such a case, this would mean
that the original data set is subject to sampling fluctuations
(sometimes called cosmic variance) which affect the clustering
measurements. Conditions 2 and 3 are related. The number of sub-samples
that should be used in the internal estimators depends on the questions
to be answered and the (unknown) form of the underlying probability
distribution: for example, to obtain an estimate of the variance
accurate to 10\% would require 50 resamplings of the data set for a
Gaussian distribution. At the same time, we need to be careful not to
make the sub-samples so small so that they become strongly correlated. 
Norberg \& Porciani (in prep.) investigate this problem using the 
two-pair correlation function. Later we address the question 
of whether or not one can put qualitative constraints, and maybe even 
quantitative constraints (using a large suite of simulations), on the 
representativeness of a given size of sub-sample. 

\subsection{Internal estimates: preliminaries} 

The first applications of non-parametric internal techniques to the estimation 
of the error on the two-point correlation function considered the removal of 
individual objects from the data set (Barrow, Bhavsar \& Sonoda 1984; 
Ling, Frenk \& Barrow 1986). This is computationally infeasible for modern 
galaxy catalogues, and it has been shown at galaxy number densities 
currently considered that this type of error estimate strongly 
underestimates the true clustering uncertainty (e.g. Mo, Jing \& B\"orner
1992; Fisher \etal\ 1994). 
Furthermore, we want to reduce the correlation between the sub-samples
into which the data set is split. For this reason, we divide the
samples by volume and split our data sets into \nsub\ cubes of equal 
volumes\footnote{We defer to another paper the discussion of the
  dependence of the results on the exact shape of the sub-volumes. For now, we
  simply note that Cabr\'e \etal\ (2007) found for angular clustering
  studies that ``irregular'' shapes can jeopardize the internal error
  methods, particularly the jackknife.}.
The internal methods then reconstruct copies of the original data sets, 
choosing or weighting the \nsub\ sub-volumes in different ways. For
each copy or resampling of the data, we make an estimate of the
correlation function, which we denote by $x^{k}_{i}$ in this
section. Here the subscript $i$ refers to the bin of spatial separation
and the superscript $k$ refers to the resampling of the data for which
the correlation function is measured. 
Note that we use the terms ``a resampling'' and ``copy'' of the data set 
interchangeably; here, a resampling refers to a ``full'' copy of a data 
set, rather than to the act of selecting an individual sub-volume. 

\subsection{Internal estimate 1: The sub-sample method} 
\label{sec:data_errsubsample}

The covariance matrix of $N$ independent realizations is, by
definition, given by
\begin{eqnarray}
C(x_i,x_j) & = & \frac{1}{N} \sum_{k=1}^{N} (x_i^k - \bar{x_i}) (x_j^k - \bar{x_j})~,
\label{eq:cov_mock}
\end{eqnarray}
where it is assumed that the mean expectation value, $\bar{x_i}$, is
not estimated from the $\{x_i^k\}_{k=1}^N$ samples, but from an
independent realization of the data.

Hence, the simplest error method, commonly referred to as the
sub-sample method, consists of splitting the data set into \nsub\ 
independent samples and estimating the covariance matrix using
Eq.~\ref{eq:cov_mock}, where the clustering statistic is estimated for
each one of the sub-samples separately. For \nsub\ independent
sub-samples, this
returns the correct covariance for a sample of volume 1/\nsub\ of the
original volume, implying that the covariance of the full dataset is
actually \nsub\ times smaller (as the variance scales directly with the
volume considered). 
This method has been used in several studies, in particular where the
survey volumes considered are large (e.g. Maddox \etal\ 1990, Hamilton
1993a and Fisher \etal\ 1994 for the APM, the IRAS 2Jy and the 
IRAS 1.2Jy galaxy surveys respectively).

However, one basic assumption made in this approach is never really
satisfied for galaxy clustering studies in the Universe: the
sub-samples are never fully independent of each other, irrespective of 
the clustering scales considered. This is due to the presence of long-range
modes in the density fluctuations, making all sub-samples to some
extent correlated with each other. This can be related to the fact that
the correlation function has a small but non-zero value on large scales. 
Therefore there is a need to consider alternative internal
estimators, accounting hopefully for these limitations. Hereafter we
will not consider the sub-sample method, even though it has been
extensively used in the past.

\subsection{Internal estimate 2: The jackknife method}
\label{sec:data_errJK}

We consider the ``delete one jackknife'' method (Shao 1986). A copy of 
the data is defined by systematically omitting, in turn, each of the 
\nsub\ sub-volumes into which the data set has been split. The
resampling of the data set consists of the $\nsubm-1$ remaining 
sub-volumes, with volume $(\nsubm-1)/\nsubm$ times the volume of 
the original data set. The clustering measurement is repeated on the
copy or resampling of the original data set. By construction, there 
are only $N=$\nsub\ different copies of the data set that are created in 
this way. The covariance matrix for $N$ jackknife resamplings is then
estimated using  
\begin{eqnarray}
C_{\rm jk}(x_i,x_j) & = & \frac{(N-1)}{N} \sum_{k=1}^{N} 
(x_i^k - \bar{x_i}) (x_j^k - \bar{x_j})~,
\label{eq:cov_jk}
\end{eqnarray}
where $x_i$ is the $i^{\rm th}$ measure of the statistic of interest 
(out of $N$ total measures), and it is assumed that the mean
expectation value is given by
\begin{eqnarray}
\bar{x_i} & = & \sum_{k=1}^N x_i^k / N \, .
\label{eq:mean_xi}
\end{eqnarray}
Note the factor of $N-1$ which appears in Eq.~\ref{eq:cov_jk}  
(Tukey 1958; Miller 1974). Qualitatively, this factor takes into account 
the lack of independence between the $N$ copies or resamplings of the data; 
recall that from one copy to the next, only two sub-volumes are
different (or equivalently $N-2$ sub-volumes are the same). Hereafter,
we will refer to a jackknife estimate from \nsub\ sub-samples as
Jack-\nsub.

\subsection{Internal estimate 3: The bootstrap method}
\label{sec:data_errBS}

A standard bootstrap resampling of the data set is made by 
selecting \nr\ sub-volumes at random, with replacement, from
the original set (Efron 1979). Effectively, a new weight is generated 
for each sub-volume. In the original data set, all sub-volumes have 
equal weight. In a resampled data set, the weight is simply the 
number of times the sub-volume has been selected e.g. 
$w_{\rm i}=0,1,2 \ldots$. The clustering measurement is repeated for 
each resampled data set. For a given \nr, the mean fractional effective 
volume\footnote{The fractional effective volume of a resampled data set
  is given by the ratio between the number of unique sub-volumes selected
  and the total number of sub-volumes the data set is split into.}
of the resampled data sets tends to a fixed fraction of the original
sample volume. For \nr=\nsub, the mean effective volume is less than
the volume of each of the jackknife resamples. We further develop this
discussion in \S\ref{sec:variance_rel} and \S\ref{sec:boot_resampling}.

Unlike for jackknife error estimates, in principle there is no limit on
the number $N$ of resamplings or copies of the data for bootstrap error
estimates. In practice, the variance on a measurement converges
relatively slowly with increasing numbers of trials (Efron \&
Tibshirani 1993). Furthermore, for our application, 
resamplings are cheap to generate but expensive to analyse. For a data
set divided into \nsub\ sub-volumes and from which one draws \nr\
sub-volumes at random with replacement, there are $(\nsubm + \nrm - 1)!/ 
(\nsubm-1)!\nrm!$ different possible bootstrap resamplings, 
which even in the modest example of \nr=\nsub=10 corresponds 
to 92~378 different bootstrap resamplings. 
Here we restrict ourselves to of the order of one hundred resamplings
(i.e. $N \sim 100$). Until now, most, if not all, bootstrap clustering
estimates have used resamplings consisting of \nr=\nsub\
sub-volumes. In this paper we test this assumption by considering
up to 4\,\nsub\ sub-volumes to construct each resampling of 
the data set (Section~\ref{sec:boot_resampling}).

The covariance matrix for $N$ bootstrap resamplings of the data is
given by 
\begin{eqnarray}
C_{\rm boot}(x_i,x_j) & = & \frac{1}{N-1} \sum_{k=1}^N (x_i^k - \bar{x_i}) (x_j^k - \bar{x_j}) , 
\label{eq:cov_boot}
\end{eqnarray}
where it is assumed that the mean expectation value is given by
Eq.~\ref{eq:mean_xi}. Note that there is no $N-1$ factor in the 
numerator of this expression, as was the case for the jackknife. 
Qualitatively the data set copies are thought of as being 
``more'' independent in the case of bootstrap resampling than 
for jackknife resampling, something we address in detail in
Sections~\ref{sec:err_anal} and~\ref{sec:test_case}. In what follows we
will refer to the mean bootstrap estimate from \nsub\ sub-samples as 
Boot-\nsub.

\subsection{External estimate: Monte Carlo realizations}
\label{sec:mock_err}

The Monte Carlo method consists of creating $N$ statistically equivalent
versions of the data set being analysed, on each of which the full
analysis is repeated. Technically, bootstrap resampling is also a Monte-Carlo 
method. However, the distinction here is that what we have termed 
the Monte Carlo approach makes no explicit reference to the observed data 
set to generate the synthetic data sets. We are not resampling the data 
in any way. Instead, we assume that we know the underlying statistical or 
physical processes which shaped the observed data set and feed these into 
a computer simulation. Here we have run N-body simulations which model  
the clustering evolution of the dark matter in the universe (see 
Section~\ref{sec:icc1340}). The Monte Carlo element in this case 
refers to the initial conditions, which are drawn from a distribution of 
density fluctuations consistent with cosmological constraints 
(see e.g. Sanchez \etal\ 2006). The level of realism 
of the computer simulation determines the cost of this method. 
For a clustering analysis of an observed galaxy catalogue, the demands 
on the Monte Carlo approach are even greater as the N-body simulations 
need to be populated with galaxies, according to some prescription, with 
the goal of statistically reproducing the galaxy sample as faithfully 
as possible (see e.g. Baugh 2008 for a review of the techniques used to
build synthetic galaxy catalogues). 

We hereafter refer to the Monte Carlo catalogues generated in the
external error estimation as ``mocks'', and their associated errors as
mock errors. In the current work using N-body simulations, we use
Eq.~\ref{eq:cov_boot} as the definition of the Monte Carlo covariance
matrix, since we define our ``reference sample'' as the mean
measurement of the correlation function extracted from the ensemble of
simulations.

\subsection{Pros and cons of each method}
\label{sec:overview}

Each of the error estimation techniques described above has its  
advantages and disadvantages. By definition, errors calculated directly
from the data take into account any hidden or unforeseen systematics
and biases that might otherwise be missed. This is particularly
important for clustering statistics, where the errors on the 2-point
correlation function also depend on the higher order clustering of the
data. These properties of the galaxy distribution are usually not 
otherwise appropriately accounted for in an error analysis. Other
properties of the data, such as the galaxy mix and survey selection,
are also naturally  satisfied in an internal approach. In the case of
external error estimation using mocks, only the statistics that have
been deliberately included in the Monte Carlo realization are
guaranteed to be taken into account in the error analysis. If a biasing
scheme has been constrained to reproduce the two-point function of
galaxy clustering, there is no guarantee that the higher order moments
of the distribution will also match those of the observed data set. 

On the other hand, internal error estimates are often severely limited 
by the size of the data set itself. This can be particularly problematic 
for clustering statistics, as studied here. Obviously, sampling or cosmic 
variance on scales larger than sample volume cannot be addressed by 
internal estimates, but can be included in external estimates made 
using mocks.

\section{The numerical experiments} 
\label{sec:err_met}

Our aim in this paper is to repeat the early comparisons of internal and 
external error estimates in the context of current and forthcoming
galaxy redshift surveys.
We draw data sets from numerical simulations which are described in 
Section~\ref{sec:icc1340}. The clustering analysis of these 
data sets is described in Section~\ref{sec:xi_2pt}. Finally, in 
Section~\ref{sec:pca}, we give an outline of the principal component 
decomposition of the covariance matrix of clustering measurements, which 
is an invaluable diagnostic in error estimation.

\subsection{Creating test data sets}
\label{sec:icc1340}

Our analysis uses the $z=0.5$ output from the L-BASICC ensemble 
of N-body simulations carried out by Angulo \etal\ (2008)\footnote{Due
  to an unfortunate labelling of the simulations outputs, we did not
  use the redshift zero outputs as initially intended.}. 
The set comprises 50 moderate resolution runs, each representing the 
dark matter using $448^{3}=89~915~392$ particles of mass 
$1.85\times10^{12}\,h^{-1}\,M_\odot$ in a box of side $1340h^{-1}$Mpc. 
Each L-BASICC run was evolved from a different realization of a Gaussian 
density field set up at $z=63$. The adopted cosmological parameters are 
broadly consistent with recent data from the cosmic microwave background 
and the power spectrum of galaxy clustering (e.g. Sanchez et~al. 2006): 
$\Omega_{\rm M}=0.25$, $\Omega_{\Lambda}=0.75$, $\sigma_{8} = 0.9$, 
$n = 1$, and $w=-1$.
Throughout we assume a value for the Hubble parameter of
$h=H_{0}/(100\,{\rm km\,s}^{-1}{\rm Mpc}^{-1})=0.73$. 

\begin{table}
  \centering
  \footnotesize
  \caption{A summary of the numerical experiments conducted in this paper. 
  Col.~1 lists the error estimation technique applied; 
  col.~2 gives the number of sub-samples into which each data set 
  is split up; 
  col.~3 indicates whether the analysis was performed in real (r) 
  or redshift (z) space; 
  col.~4 gives the number of different data sets used; 
  col.~5 shows the number of resamplings and clustering measurements 
  performed for each data set, $N$; 
  col.~6 lists, for the case of bootstrap errors only, the relative
  number of sub-volumes selected at random with replacement from the
  original list w.r.t. \nsub;  
  col.~7 shows the sampling fraction w.r.t. our nominal mean density
  which is set to match that of a \lstar\ galaxy sample. 
  The first group of experiments yielded our main results 
  and used in the majority of the plots; the other experiments are 
  variants to test different components of the analysis and referred 
  to in the text.  
  }
  \label{tab:sum_runs}
  \begin{tabular}{ccccccc}
    \hline \hline
    {Error} & \nsub\ & r/z & $N_{data set}$ & 
      $N$ & \nr/\nsub\ & f\\
    \hline
      Mock &   1 & z [r] & 100 [100] &  1 &   & 1\\
      Boot &   8 & z [r] & 100 [50]  & 99 & 1 & 1\\
      Boot &  27 & z [r] & 100 [50]  & 99 & 1 & 1 \\
      Jack &   8 & z [r] & 100 [50]  &  8 &   & 1\\
      Jack &  27 & z [r] & 100 [50]  & 27 &   & 1 \\
      Jack &  64 & z     & 100       & 64 &   & 1\\
    \hline
      Mock &   1 & r & 100 [100] &   1 &      & 0.1 [0.25]\\
      Boot &   8 & r &  50 [50]  &  99 & 1    & 0.1 [0.25]\\
      Boot &  27 & r &  50 [50]  &  99 & 1    & 0.1 [0.25]\\
      Boot &  64 & r &  50 [50]  &  99 & 1    & 0.1 [0.25]\\
      Boot & 125 & r &  50 [50]  & 199 & 1    & 0.1 [0.25]\\
      Boot &   8 & r &  25 [25]  &  99 & 2, 3 & 0.1 [0.25]\\
      Boot &  27 & r &  25 [100] &  99 & 2, 3 & 0.1 [0.25]\\
      Boot &  64 & r &  25 [100] &  99 & 2, 3 & 0.1 [0.25]\\
      Boot & 125 & r &  25 [100] & 199 & 2, 3 & 0.1 [0.25]\\
      Jack &   8 & r & 100 [50]  &   8 &      & 0.1 [0.25]\\
      Jack &  27 & r & 100 [100] &  27 &      & 0.1 [0.25]\\
      Jack &  64 & r & 100 [100] &  64 &      & 0.1 [0.25]\\
      Jack & 125 & r & 100 [100] & 125 &      & 0.1 [0.25]\\
    \hline
      Boot &   8 & r & 6 [6] &  99 & 4 & 0.1 [0.25]\\
      Boot &  27 & r & 6 [6] &  99 & 4 & 0.1 [0.25]\\
      Boot &  64 & r & 6 [6] &  99 & 4 & 0.1 [0.25]\\
      Boot & 125 & r & 6 [6] & 199 & 4 & 0.1 [0.25]\\
    \hline \hline
  \end{tabular}
\end{table}

The combination of a large number of independent realizations and the
huge simulation volume make the L-BASICC ensemble ideal for our
purposes. Angulo et~al. (2008) showed that nonlinear and dynamical
effects are still important for the evolution of matter fluctuations 
even on scales in excess of $100 h^{-1}$Mpc and that these scales 
contribute to the growth of smaller scale perturbations. A smaller
simulation volume would not be able to model the growth of fluctuations
accurately because these long wavelength modes would be missed. The
large volume also means that it is still possible to extract more than 
one region from each simulation which can be considered as being
effectively independent from the others due to their large spatial
separation. Equally important it is also possible to mimic with such a
large simulation volume the real situation of a survey like the 2dFGRS,
for which there are two distinct survey regions in two nearly opposite
directions on the sky. 

From each of the 50 simulation cubes we extract two cubical sub-volumes 
of $380~\Mpc$ on a side which are separated by at least $\sim 500~\Mpc$ 
from one other. We could have extracted more than 40 volumes of 
this size from each simulation cube without choosing the same volume 
twice. However, we chose to follow this more conservative approach 
as we want to be able to treat the two sub-volumes as being effectively 
independent of one another. Although they come from the same simulation 
volume, the fluctuations on scales greater than $500~\Mpc$ are relatively 
weak. Hence, the total number of available data sets constructed in this 
way comes to 100, each of which is fully independent of 98 of the others 
and essentially independent of the remaining 99$^{\rm th}$ data
set. The size of each data set is chosen to match the volume of a
typical \lstar\ volume-limited sample extracted from the main SDSS
galaxy redshift survey (e.g. Zehavi \etal\ 2004, 2005). Note that this
volume is about 10 times larger than the equivalent volume-limited
samples of \lstar\ galaxies used in the final 2dFGRS clustering
analyses (Norberg \etal\ in prep.). Finally,
we have randomly diluted the number of dark matter particles in each 
data set in order to match the number density of a \lstar\ galaxy 
sample of $3.7 \times 10^{-3}~\Mpccube$, which mimics the discreteness or 
shot noise level in typical observational data sets. 

Note that on this occasion, we do not attempt to model a particular galaxy 
sample in detail, as our aim here is to provide a generic analysis that is 
as widely applicable and reproducible as possible, within a framework which 
is likely to remain the same in the foreseeable future. Moreover, if one 
attempted to model a galaxy sample rather than the dark matter, this would 
open up the issue of how well can the model reproduce the higher order 
clustering of the galaxies. This is a problem to be addressed by those 
building mock galaxy catalogues for particular surveys and is beyond the 
scope of the current paper. By focusing on dark matter clustering only, 
our analysis remains well defined and fully accurate within our chosen 
cosmological model.

A summary of the samples and error calculations carried out is given in 
Table~\ref{tab:sum_runs}. Each data set drawn from a simulation can be 
divided into sub-volumes in order that it be resampled according to the 
jackknife or bootstrap algorithm. Thus, we have up to 100 independent 
experiments for which we can compare internal estimates of the error 
on the correlation function, and consider for the first time in detail
the whole error distribution of internal error estimates. The external 
``mock'' estimate of the error is obtained using all 100 data sets. In
all, we have performed about 250 thousand correlation function
estimates, representing about one million CPU hours of
calculations. 

\subsection{Estimating the two-point correlation function}
\label{sec:xi_2pt}

Throughout this paper we consider only standard two-point clustering
statistics and we delay to a future paper the study of more recent
two-point clustering statistics, like the $\omega$-statistic of
Padmanabhan, White \& Eisenstein (2007).
We measure the two point correlation function, 
$\xi_{\rm X}(r_{\rm p},\pi)$, as a function of pair separation
perpendicular to ($r_{\rm p}$) and parallel to ($\pi$) the line of
sight of a fiducial observer, in real (X=$r$) and in redshift space
(X=$s$). In order to remain as comparable as possible with the analysis 
of observational data, we do {\it not} use the common distant observer 
approximation, but instead perform a realistic angular treatment with 
a location chosen for the observer within the simulation box, i.e.: 
\begin{eqnarray}
{\bf r_{los}} & = & ({\bf \widehat{r_1}} + {\bf \widehat{r_2}})/2  \nonumber \\ 
r_p & = & {\bf r_1} \cdot {\bf \widehat{r_{los}}} + {\bf r_2} \cdot {\bf
  \widehat{r_{los}}}  \nonumber \\ 
\pi & = & \sqrt{({\bf r_1}-{\bf r_2})^2-(r_p)^2}~,
\end{eqnarray}
where ${\bf r_1}$ and ${\bf r_2}$ are the position vectors of the two objects 
in the pair and a hat indicates that the vector is appropriately normalized; 
${\bf r_{los}}$ is the mean distance to the pair of objects along the 
line of sight from the observer. 
Redshift space distortions are modelled in the same way, i.e.: 
\begin{eqnarray}
z_{\rm tot} & = & z_{\rm hub} + {\bf v} \cdot {\bf \widehat{r}} / c 
\,,
\end{eqnarray}
where $z_{\rm hub}$ is the galaxy's redshift without peculiar motions, 
${\bf v}$ is its peculiar velocity in addition to the Hubble flow 
and $c$ is the speed of light. 

We calculate the two dimensional correlation function 
$\xi_{\rm X}(r_{\rm p},\pi)$ 
for each data set using standard estimators, e.g. 
Hamilton (1993) and Landy \& Szalay (1993). In order to obtain a robust 
estimate of the mean density, these estimators require that a large number 
of points be randomly distributed within the boundaries of the data set, 
according to the angular and radial selection functions which define the 
data. The number of randomly distributed points is typically larger than 
the number of data points by a factor of $f \simeq 100$, and we use the
technique outlined in Landy \& Szalay (1993) to speed up the
calculation of the random-random pair counts. 
In practice, to avoid fluctuations in the mean density at small pair
separations, we typically use 4 times as many random points as data
points and repeat the estimate of the pair counts about 25 times. 
We use a unique set of randoms for each calculation, so that
our statistics are not influenced by any features in the
randoms, which when an experiment is repeated numerous times could
actually show up in a systematic fashion. 
Finally to compute the $\xi_{\rm s}(r_{\rm p},\pi)$ estimate, we limit the
counts to pairs separated by less than 10 degrees for two reasons: 
1) results effectively estimated in redshift space can be properly
interpreted using the redshift space distortion model of Kaiser (1987),
developed in the distant observer approximation; 2) the projected
correlation function, defined below, can be considered as a real
space quantity. These issues are discussed in greater detail in e.g.
Matsubara (2000) and Szapudi (2004), but also applied in some
clustering studies, like Hawkins \etal\ (2003) and 
Cabr\'e \& Gazta\~naga (2008). 
%
Using logarithmic binning in both the $r_{\rm p}$ and $\pi$-directions, we
estimate the projected correlation function, \wprp \, (sometimes also written 
as $\Xi(\sigma)/\sigma$ and with $\sigma=r_{\rm p}$), by 
integrating $\xi_{\rm X}(r_{\rm p},\pi)$ in the $\pi$ direction 
(see Eq.~\ref{eq:wp_rp} below). The only real motivation for this 
exercise is that the projected correlation function is, in theory, 
free from distortions arising from gravitationally induced peculiar 
motions. This point is discussed further in Section~\ref{sec:sum_runs}.

We also estimate the spherically averaged correlation
functions, $\xi_r(s)$ and $\xi_s(s)$ in ``$r$'' real 
and ``$s$'' redshift space respectively, by averaging over shells 
in pair separation given by $s = \sqrt{r^{2}_{p} + \pi^2}$. As 
for the projected correlation function we use logarithmic binning 
and accumulate pair counts directly, i.e. not by integrating over 
the 2-d correlation function estimate.

Given $\xi_{\rm X}(r_{\rm p},\pi)$, the different clustering statistics are
related by 
\begin{eqnarray}
\label{eq:wp_rp}
w_{\rm p}(r_{\rm p})/r_{\rm p} & = & \frac{2}{r_{\rm p}} \int_{0}^{\pi_{\rm max}}
\xi_{\rm X}(r_{\rm p},\pi)\, d\pi ~.  \\
\xi_{\rm X}(s) & = & < \xi_{\rm X}(r_{\rm p},\pi) >|_{s = \sqrt{r^{2}_{\rm p} + \pi^2} } \, ,
\label{eq:xi_proj}
\end{eqnarray}
with X=$r$ or $s$, real and redshift space respectively. Here, the integral 
for $w_{\rm p}(r_{\rm p})/r_{\rm p}$ is carried out to a maximum value of 
the separation along the line of sight of $\pi_{\rm max}$. 
Theoretically \wp\ is only a true real space quantity 
when $\pi_{max}=\infty$. In Section~\ref{sec:sum_runs} we discuss the 
systematic implications of a finite choice for $\pi_{\rm max}$.

\subsection{Principal component decomposition}
\label{sec:pca}

The covariance matrix of correlation function measurements, is not, 
in itself, of much interest. However, the inverse of the covariance 
matrix plays a pivotal role, appearing, for example, in all simple 
$\chi^2$ estimates. 
As matrix inversions are highly non-linear by nature, the impact of 
noise in the estimate of the covariance matrix is hard to judge, 
unless extensive tests are carried out. A number of procedures exist 
to help with the matrix inversion and analysis.  

First, we prewhiten 
the covariance matrix, which means that all diagonal terms are rescaled 
to unity and all non-diagonal terms are rescaled to fall between -1 and 1, 
\begin{eqnarray}
{\bf C} & = & {\bf \sigma} \, {\bf C_{N}} \, {\bf \sigma}~,
\label{eq:cov_norm}
\end{eqnarray}
where ${\bf C_{N}}$ is the normalized covariance matrix, and
$\sigma_{i,j} = C_{i,i} \delta_{i,j}$, hereafter also referred 
to as $\sigma_{i}$. 

Second, we decompose the $N_u \times N_u$ normalized covariance matrix
into its principal components, solving the eigen-equations: 
\begin{eqnarray}
{\bf C_{N}} \, {\bf E_i} & = & \lambda_i \, {\bf E_i}  \quad i=1,...,N_u~,
\label{eq:eigen}
\end{eqnarray}
where $\lambda_i$ and ${\bf E_i}$ are the normalized eigenvalues and
eigenvectors of the normalized covariance matrix. We have deliberately
defined $N_u$ as the dimension of the covariance matrix actually 
used in the analysis, and not assumed it to be necessarily equal to the 
total number of data points for the given statistic (e.g. the number of bins 
of pair separation in the correlation function). Even though the parts of
the covariance matrix that are used do not depend on those that are not
used, principal component decomposition and matrix inversion strongly
rely on the full matrix set up.
For that reason, it is essential to pre-define the parts of the
covariance matrix to be considered, a fact that is most often
neglected.

There are many useful properties of principal component decomposition worth 
listing here (see for example Kendall 1975; Press et~al. 1992): 
\begin{itemize}
\item The $\left\{{\bf E_i}\right\}_{i=1}^{N}$ form a complete orthogonal
  basis, from which the whole covariance matrix can be reconstructed.   
\item A principal component decomposition minimises the error for the 
  number of components used. 
\item A ranking of the eigenvectors in order of decreasing eigenvalue
  highlights the principal trends of the covariance matrix. 
\item For a given covariance matrix there is a unique set of principal
  components. Comparing covariance matrices is equivalent to the
  comparison of principal components. 
\end{itemize}
With respect to this last point, it is essential to understand noise in
the statistic and its non-trivial effect on estimating the principal
components of the covariance matrix. Clearly, in the absence of noise,
two covariance matrices are identical if, and only if, their principal
components are identical. In real systems that contain noise this is no
longer true.

For this paper the most important use of the principal component
decomposition technique is to redefine the $\chi^2$ of a model given
the data: 
\begin{eqnarray}
\chi^2 & = & ({\bf x_d} - {\bf x_{th}})^T \, {\bf C}^{-1} \, ({\bf x_d} - {\bf x_{th}}) 
\label{eq:chi2}\\
 & = & \sum_{i=1}^{N_u} \frac{({\bf y_d}_i - {\bf y_{th}}_i)^2}{\lambda_i} ~,
\label{eq:chi2_pca} \\
{\bf y_d} & = & {\bf E}^T \, {\bf \sigma}^{-1} \, {\bf x_d} ~,
\end{eqnarray}
with ${\bf \sigma}$ defined by Eq.~\ref{eq:cov_norm}, and where ${\bf E}$ 
is the rotation matrix composed of the unique eigenvectors ${\bf E_i}$
previously defined in Eq.~\ref{eq:eigen}. The beauty of 
Eq.~\ref{eq:chi2_pca} is that, when summed over all $N_u$ terms, it is
exactly equivalent to the standard $\chi^2$ given in
Eq.~\ref{eq:chi2}. Additionally, the properties of the principal
component decomposition ensure it yields the most efficient
representation of the covariance matrix if we truncate it to include
fewer than $N_u$ modes. Hence, hereafter, the number of principal
components considered is given by N$_{\rm pca}$, and only when 
N$_{\rm pca} = N_u$ are we considering a full covariance matrix
analysis. 

\begin{figure*}
\plottwo{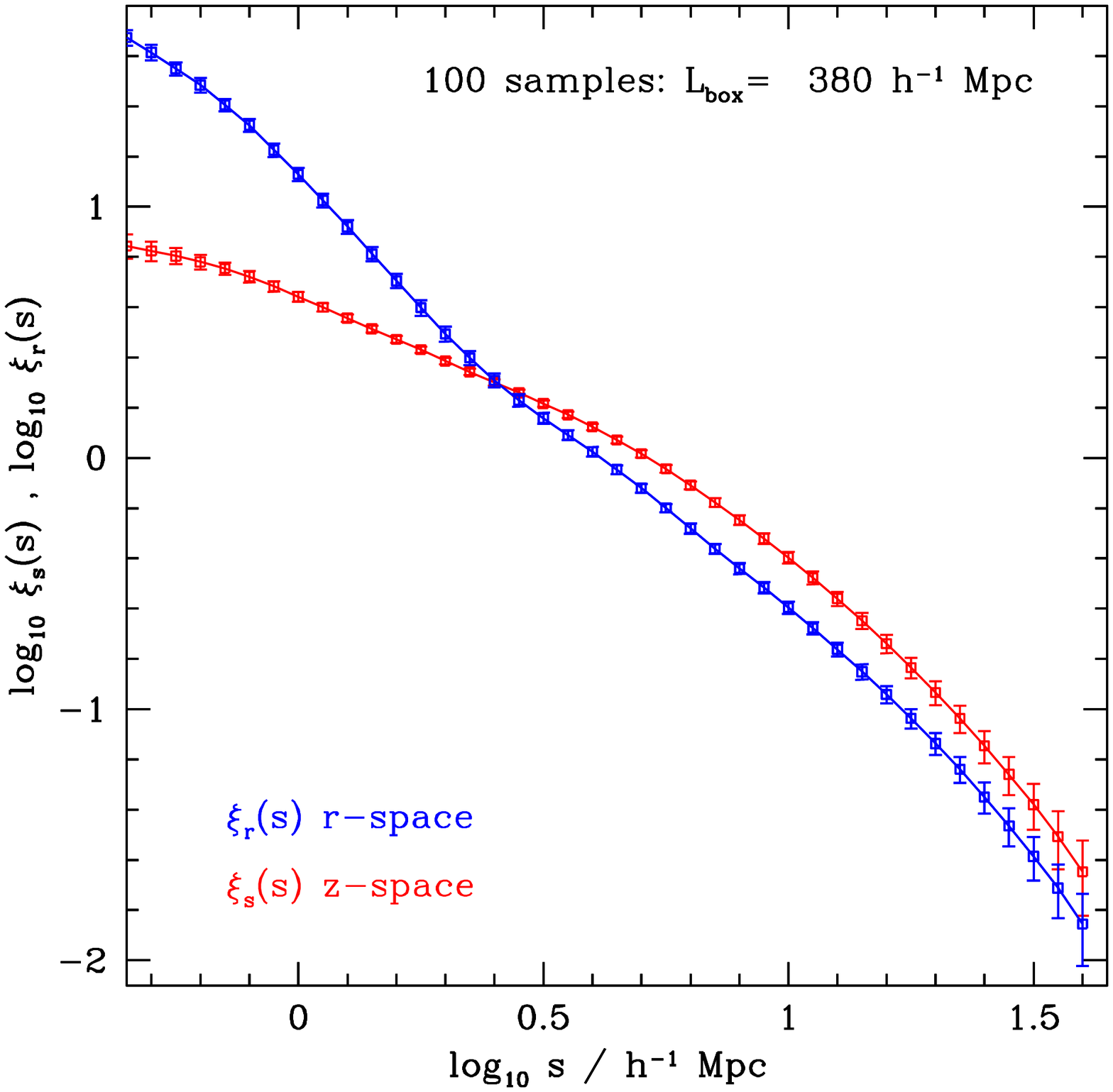}{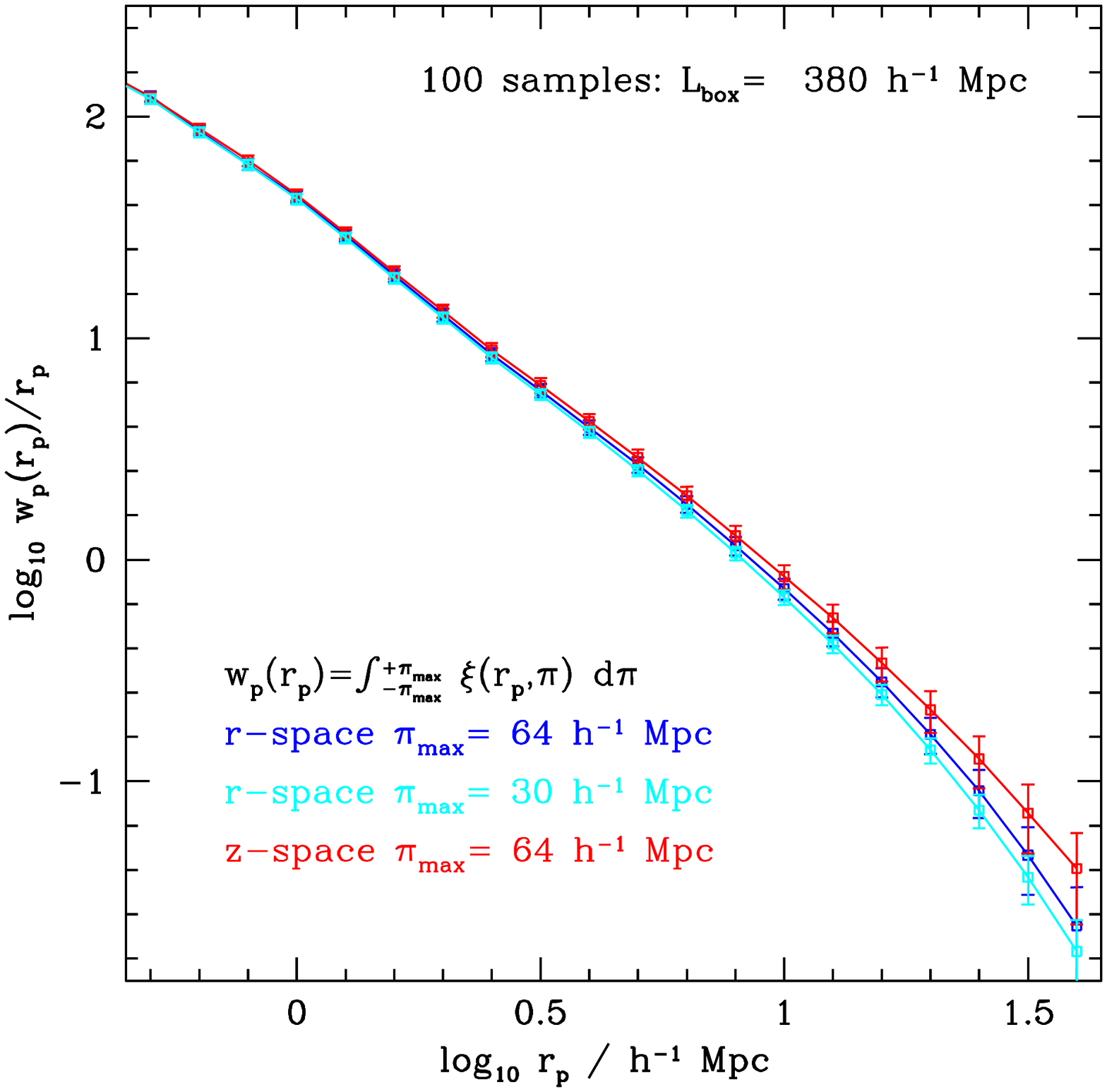}
\caption
{
{\it Left:}
The spherically averaged correlation function of the 
dark matter, $\xi(s)$, estimated in real (blue) and redshift space (red)
for 100 data sets extracted from the L-BASICC simulation ensemble, as 
described in \S\ref{sec:icc1340}.
{\it Right:}
The projected correlation function, \wp, of the dark matter 
estimated directly in real space and in redshift space by integrating 
$\xi_{\rm X}(r_{\rm p}, \pi)$ for the same data sets as used in the
left panel. 
The projected correlation functions are integrated out to two limits:
$\pi_{\rm max}=64$ \& $30\Mpc$, as indicated by the legend. 
For clarity, the redshift space estimate integrated out 
to $\pi_{\rm max}=30\Mpc$ is omitted. 
In both panels, the error on the mean of the quantity is plotted. See
\S\ref{sec:sum_runs} for comments.
}
\label{fig:clustering}
\end{figure*}

\section{Analysis and Results} 
\label{sec:err_anal}

In this section we present our main results, carrying out a systematic 
comparison of applying different error estimation techniques to data sets 
which are representative of current and forthcoming galaxy surveys. 
The numerical experiments we have carried out are summarised in 
Table~\ref{tab:sum_runs}. To recap, we have created 100 test data sets by 
extracting independent volumes from an ensemble of N-body simulations
(\S\ref{sec:icc1340}).  
In our fiducial case, the density of dark matter particles in the test 
data sets has been diluted to match the abundance of \lstar\ galaxies. 
Each data set can be divided into equal sized cubical sub-volumes and 
resampled according to the jackknife and bootstrap algorithms to make 
internal estimates of the 
error on the two-point correlation function. We examine the scatter in the 
internal error estimates by comparing results obtained from different 
data sets. In our analysis, we vary the number of sub-samples each data 
set is divided up into, the sampling rate of the particles, the number 
of sub-volumes selected to construct a copy of the data set in the 
case of the bootstrap and also show results for clustering measured 
in real and redshift-space. Our benchmark in this study is the external 
estimate of the error obtained by treating our data sets as 100 independent 
experiments. This is regarded as the ``truth" in our paper. 

This section is quite long and rather detailed. For that reason, on a
first reading of the paper the reader may wish to go directly to the
summary in Section~\ref{sec:conclusion} or to the case study in
Section~\ref{sec:test_case}. In Section~\ref{sec:sum_runs} we  
present the correlation function measurements made from the ensemble of 
data sets. We then go through different aspects of the error analysis: 
in Section~\ref{sec:variance_rel} we compare the relative errors obtained 
from the different techniques; in Section~\ref{sec:error_dist} we look at 
the uncertainty in the error; in Section~\ref{sec:eigen_val} and 
Section~\ref{sec:eigen_vec} we examine the distribution of eigenvalues
and eigenvectors respectively of the covariance matrices produced by
each approach, before ending in Section~\ref{sec:stab_covmat} with a
discussion of the stability of the inversion of the respective
covariance matrices.

\subsection{Correlation functions}
\label{sec:sum_runs}

We present results for the projected correlation function and
the spherically averaged correlation functions in both real and redshift
space. Selected results later on in the Section depend on the 
number of bins used to represent the correlation function. 
In most plots we focus on the scales 
$1.0 \lesssim r / \Mpc \lesssim 25$ for the following reasons. First,
this range is accurately modelled in the L-BASICC simulations, and can
be measured reliably within all of the sub-volumes the data sets are
split into. 
Second, the model fitting presented in Section~\ref{sec:test_case} 
is only valid down to $\sim 1.0~\Mpc$, and it therefore makes sense to present 
other results over the same range of scales for comparison. Third, this 
range is the most appropriate to draw direct analogies with more general 
analyses of the error in galaxy clustering measurements.  On sub-Mpc 
scales the precise way that galaxies are distributed 
within their host halo strongly influences the
correlation function and its error. In real-space, scales larger 
than $\sim 20~\Mpc$ are even more problematic as estimates can be 
affected by systematics at the $\sim 30$~\% level 
(see Fig.~\ref{fig:clustering}), if special care is not taken when
accounting for redshift space distortions. Fourth, this choice of scale
produces clustering estimates that can be split into sufficiently small
but still well sampled bins.

We first present measurements of the basic correlation functions 
we will be working with throughout the remainder of the paper. 
Fig.~\ref{fig:clustering} shows the spherically averaged correlation 
functions, $\xi_r(s)$ and $\xi_s(s)$, and the projected correlation 
function, \wprp, in the left and right panels respectively, as 
estimated from the 100 independent data sets of $380~\Mpc$ aside. 
In the left panel of Fig.~\ref{fig:clustering}, we see the clear 
impact of redshift space distortions on the measured $\xi_s(s)$.  
On small scales, the clustering signal is severely damped by the 
peculiar motions of the dark matter particles (responsible for the 
fingers-of-God due to cluster-mass haloes in a $\xi_s(r_{\rm p},\pi)$ 
representation). On intermediate scales, large scale infall enhances 
the clustering amplitude compared with the real space measurement 
(see Peacock \& Dodds 1994 for a model of the small and large scale
redshift space distortion). 
$\xi_r(s)$ shows the well known shoulder between 1 and $3~\Mpc$ 
(Zehavi \etal\ 2004), a sign of the transition between the one and 
two halo terms used in halo occupation distribution models (e.g. Berlind 
\etal\ 2003). 

\begin{figure*}
\plotthree{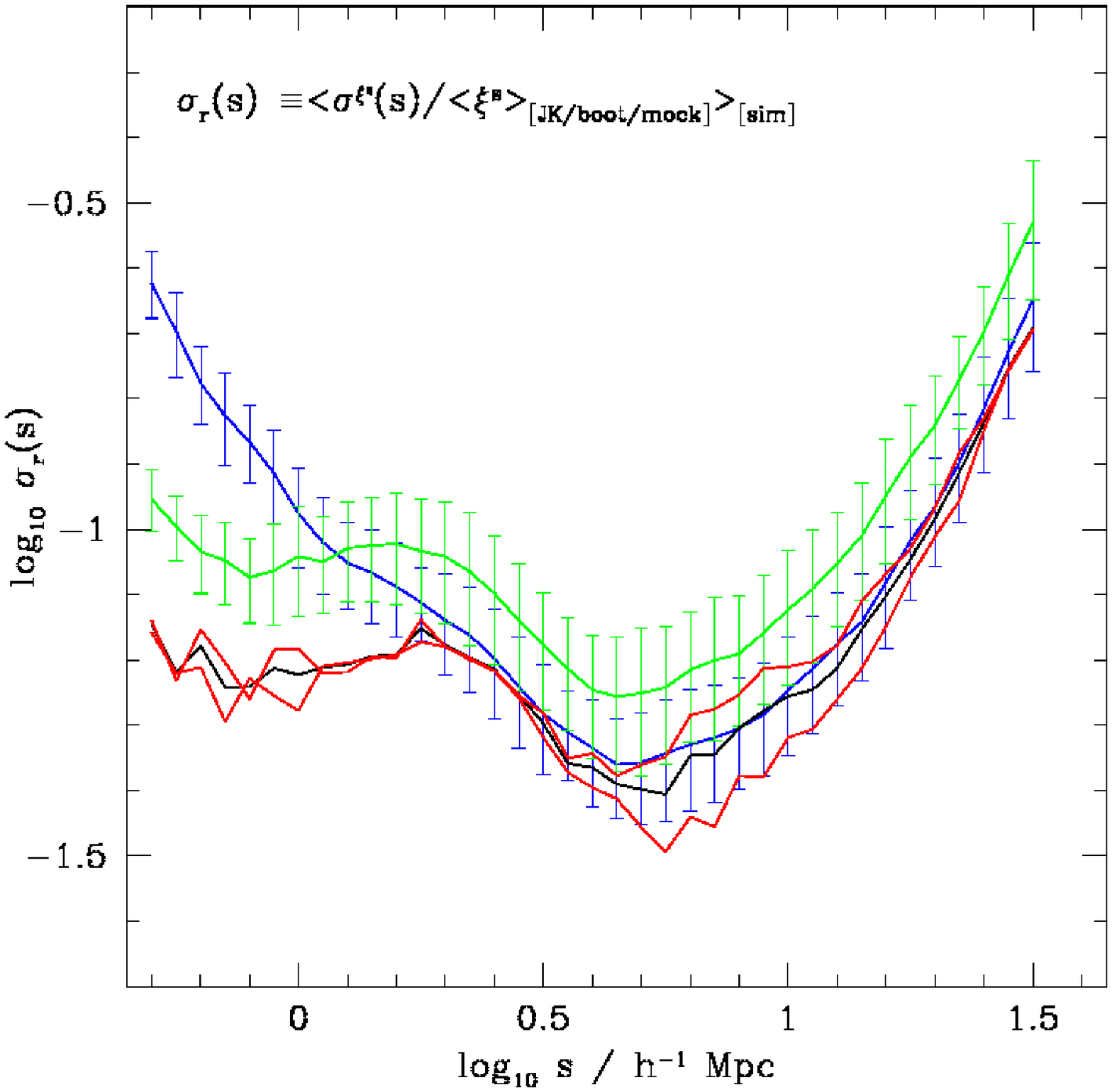}{0.33}{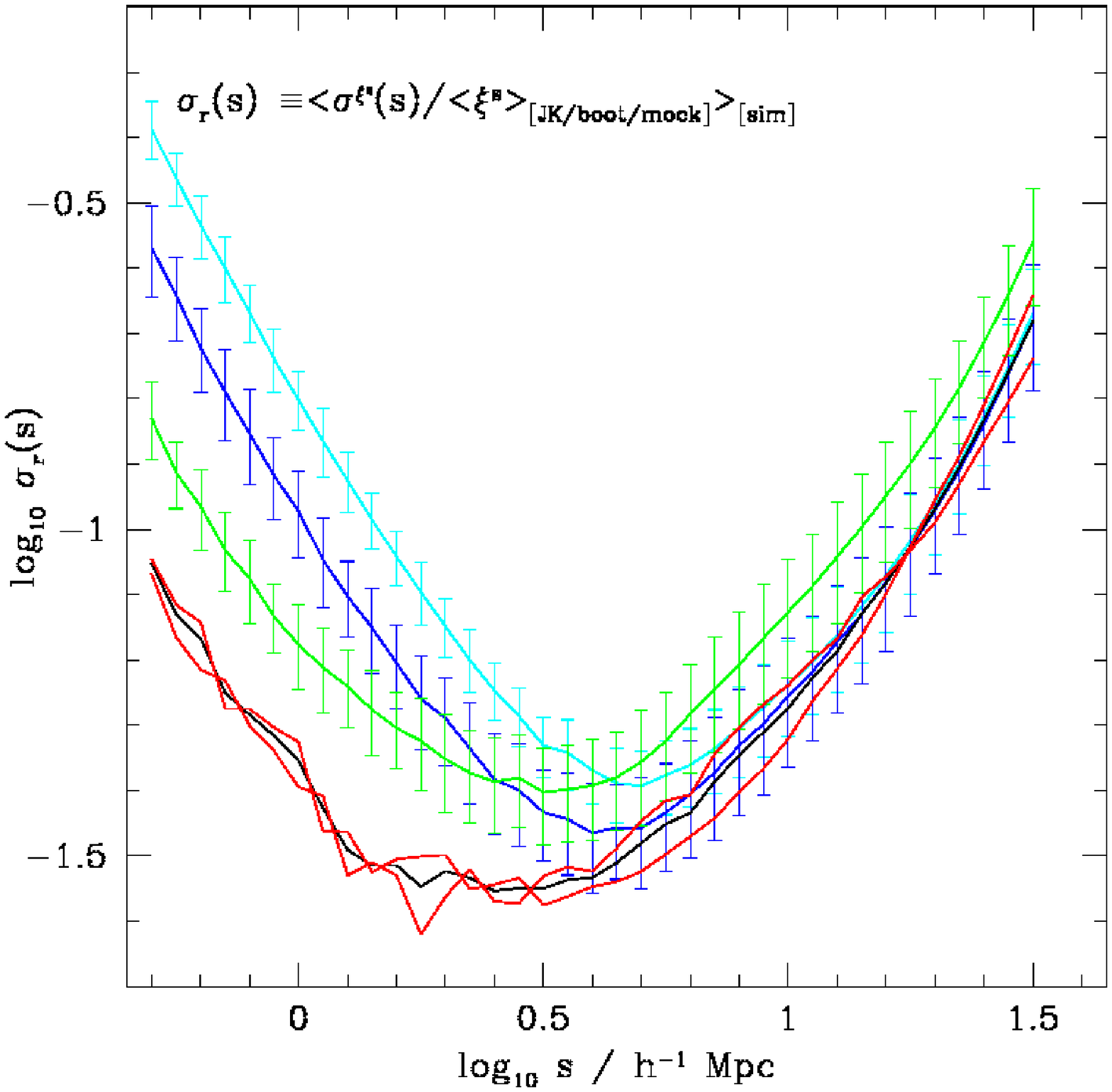}{0.33}{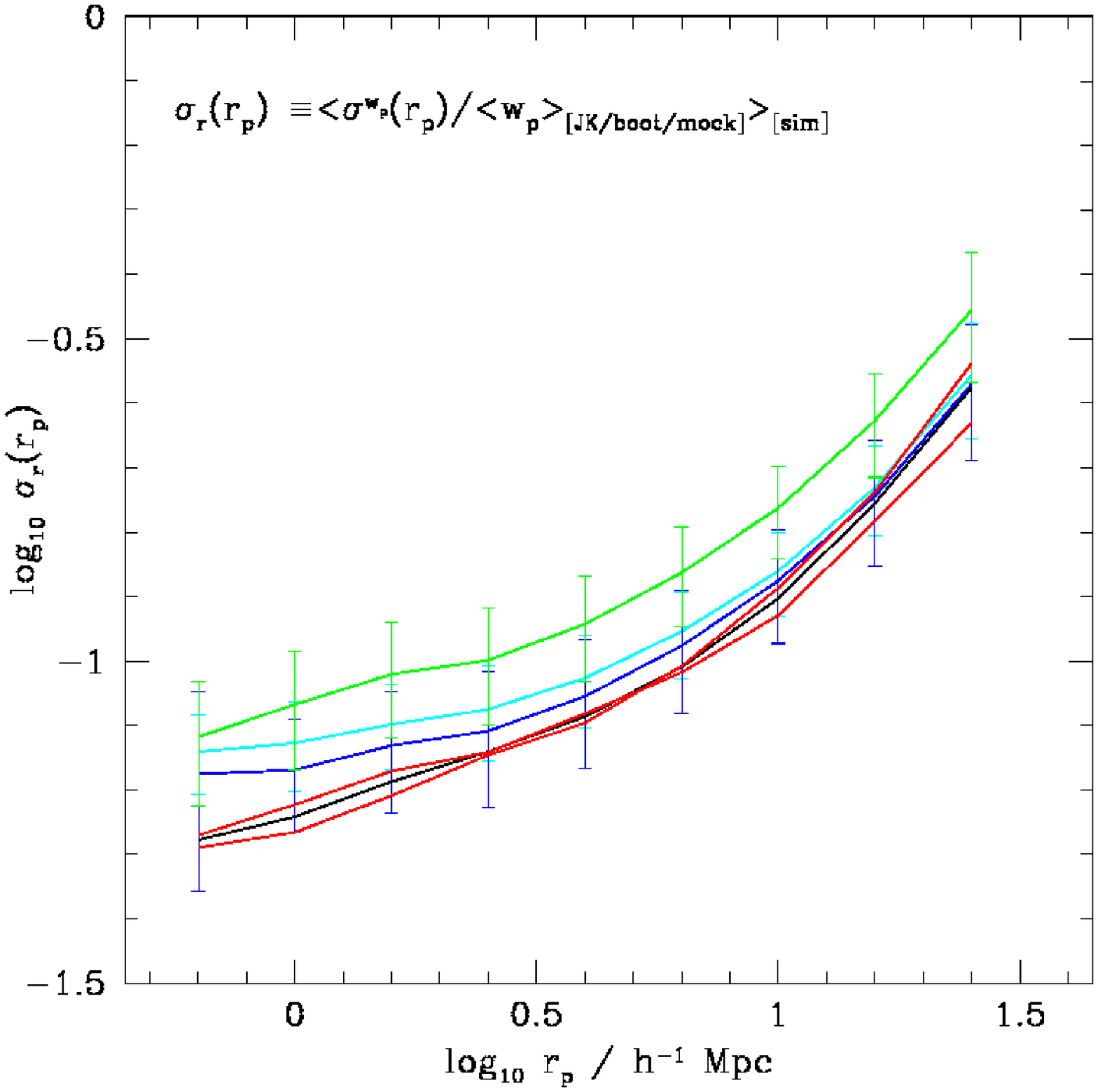}{0.33}
\caption
{
The mean relative variance of $\xi_{r}(s)$ (left), $\xi_{s}(s)$ (middle) 
and \wp\ (right) as a function of scale for different error estimators: 
black -- 100 ``mock'' data sets; red -- two different samples of the mocks 
(the first 50 \& the last 50); green -- bootstrap errors after splitting 
each data set into 27 sub-samples (i.e. Boot-27 with \nr=\nsub); 
blue and cyan -- jackknife errors measured after splitting each data set into 
27 and 64 sub-samples respectively (i.e. Jack-27 and Jack-64). See
\S\ref{sec:variance_rel} for comments.
}
\label{fig:variance}
\end{figure*}

In the right hand panel of Fig.~\ref{fig:clustering} we show the
projected correlation functions, \wprp, in real and redshift space. 
By real-space, here we mean that we integrate over an estimate of 
$\xi_r(r_{\rm p}, \pi)$ made without including peculiar motions. 
The comparison between these results illustrates the 
sensitivity of the projected correlation function to the 
choice of $\pi_{\rm max}$ in Eq.~\ref{eq:xi_proj}. Theoretically, 
if one integrates out to $\pi_{\rm max}=\infty$, Eq.~\ref{eq:xi_proj} 
would return a purely real space quantity.  For numerical as well 
as computational reasons\footnote{
  The larger the integration range, the more computational 
  resources are needed to perform the pair count to estimate 
  $\xi_{\rm X}(r_{\rm p}, \pi)$, since the number of pairs increases in 
  proportion to the additional volume included as $\pi$ increases.} 
we restrict the integration to $\pi_{\rm max} \leq 64~\Mpc$. 
Even at $r_{\rm p} \sim 10~\Mpc$ there is a systematic difference 
of 10~\% in this case between \wprp\ estimated for data in real 
and redshift space. 
This difference increases with scale, and by $\sim 30~\Mpc$ it is 
greater than 50~\%. Due to the large sub-volumes used this systematic 
effect is statistically non-negligible. Taking the individual 
errorbars at face value (which we show below to be ill-advised), at
$\sim 20~\Mpc$ this difference is about 1-$\sigma$. For comparison
purposes, we plot \wprp\ estimated in real 
space data using $\pi_{\rm max}=30~\Mpc$ (cyan). Even in real space, 
the chosen value of $\pi_{\rm max}$ has non-negligible consequences. 

The fact that the real and redshift space measurements of the projected 
correlation function do not perfectly agree in the right hand panel of 
Fig.~\ref{fig:clustering} has important implications for the interpretation 
of \wp. An estimate of the projected correlation function made 
from data in redshift space will still be influenced by the underlying 
redshift space distortions if the integral is not performed out to large 
enough scales. In our analysis, the magnitude 
of the systematic shifts is calculated for simulated cold dark matter 
particles. In the case of real galaxies, some of which are strongly 
biased with respect to the underlying dark matter, the discrepancy 
could be even larger. One should therefore be cautious about considering 
\wp\ as a ``real space'' measurement, and take even more care in
understanding the effect of the limiting $\pi_{\rm max}$ value.

\subsection{Mean relative variance as function of scale}
\label{sec:variance_rel}

In Fig.~\ref{fig:variance} we present the mean relative variance of
$\xi_r(s)$ (left), $\xi_s(s)$ (centre) and \wp\ (right), all as
function of scale for different error estimators. The mean relative 
variance for each statistic is defined in the top left corner of 
each panel.

The relative variance on the clustering measurements estimated from 
the scatter over the 100 data sets is robust (solid black). If we
consider the first 50 and last 50 data sets as separate ensembles
(solid red), we find very similar 
results for the relative variance. However, for all three statistics 
considered ($\xi_r(s)$, $\xi_s(s)$ and \wp), neither the bootstrap 
nor jackknife is able to fully recover the relative variance of the mocks 
as function of scale. Bootstrap estimates for which the number of sub-volumes 
selected, \nr, is equal to \nsub\ tend to systematically overestimate 
the relative variance by 40-50~\% on all scales. Jackknife estimates 
show a scale and statistic dependent bias, varying from no bias on scales 
larger than $\sim 10~\Mpc$, to a 25~\% overestimate for \wp, and as much 
as 200 to 400~\% for $\xi_r(s)$ and $\xi_s(s)$ on sub-Mpc scales. We have 
checked that these differences are not due to a systematic uncertainty in 
the recovered mean of each statistic: even when rescaling the variance by 
a simple power-law, the conclusions remain unchanged.

\begin{figure*}
\plotthree{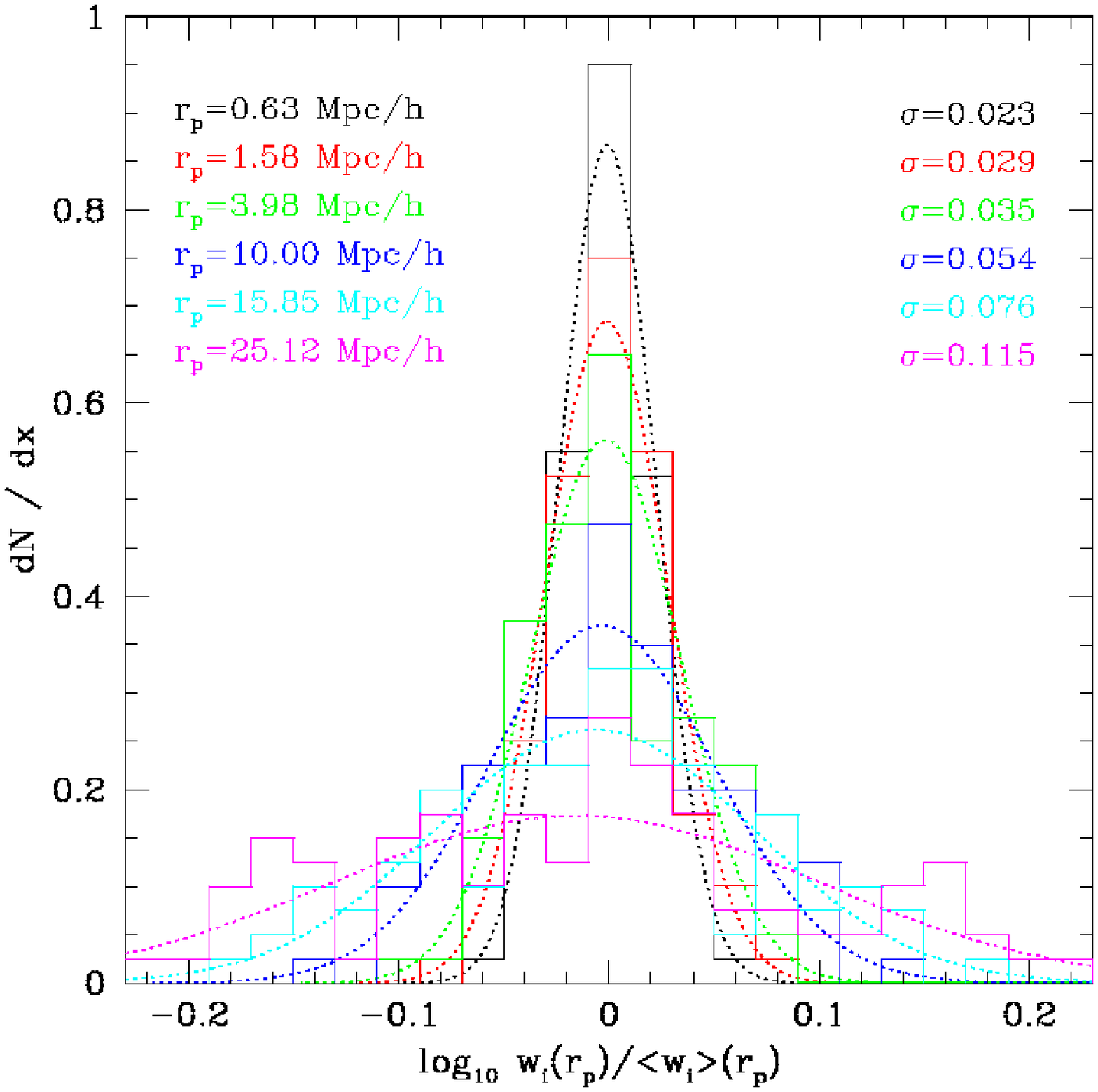}{0.33}{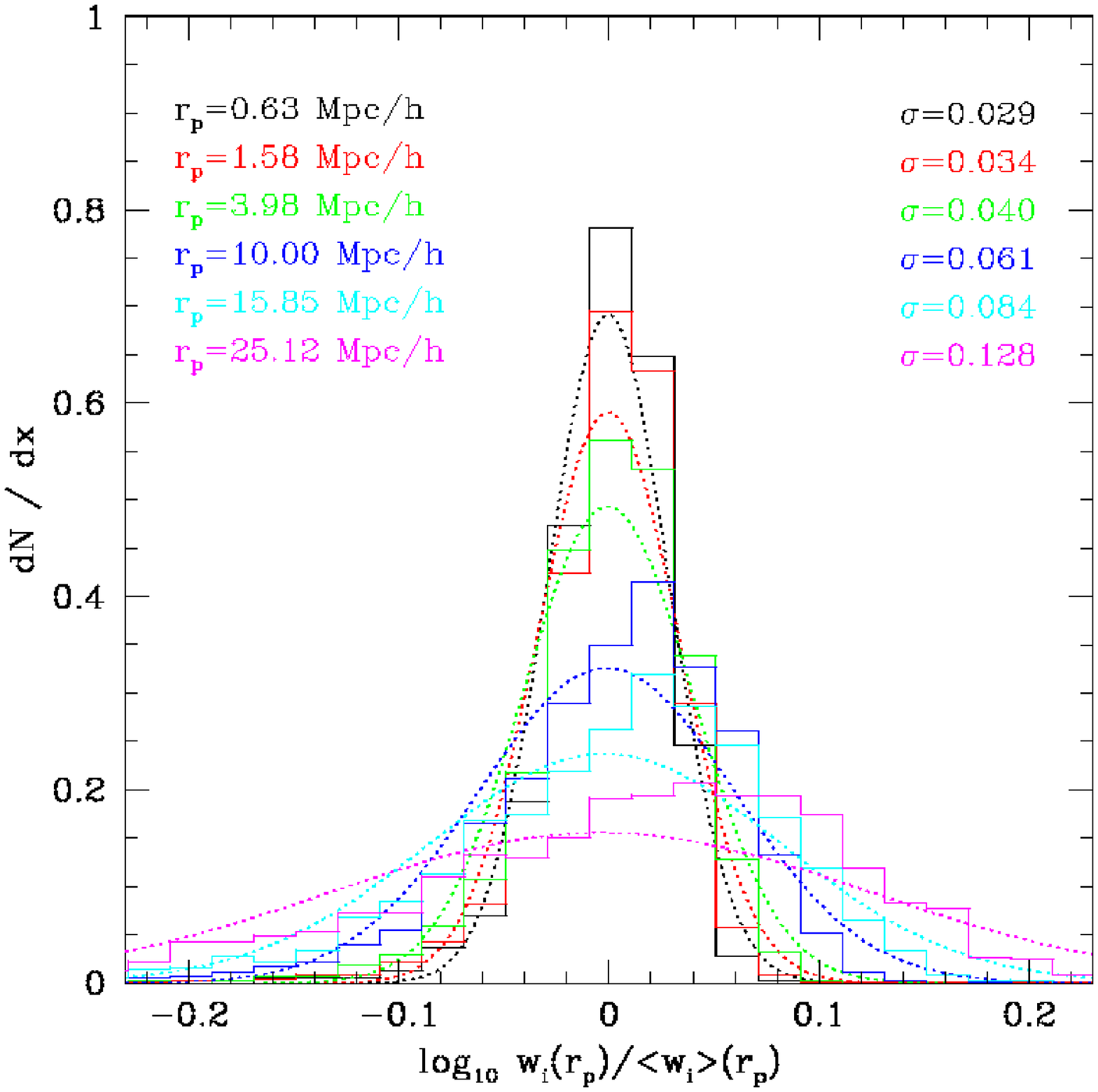}{0.33}{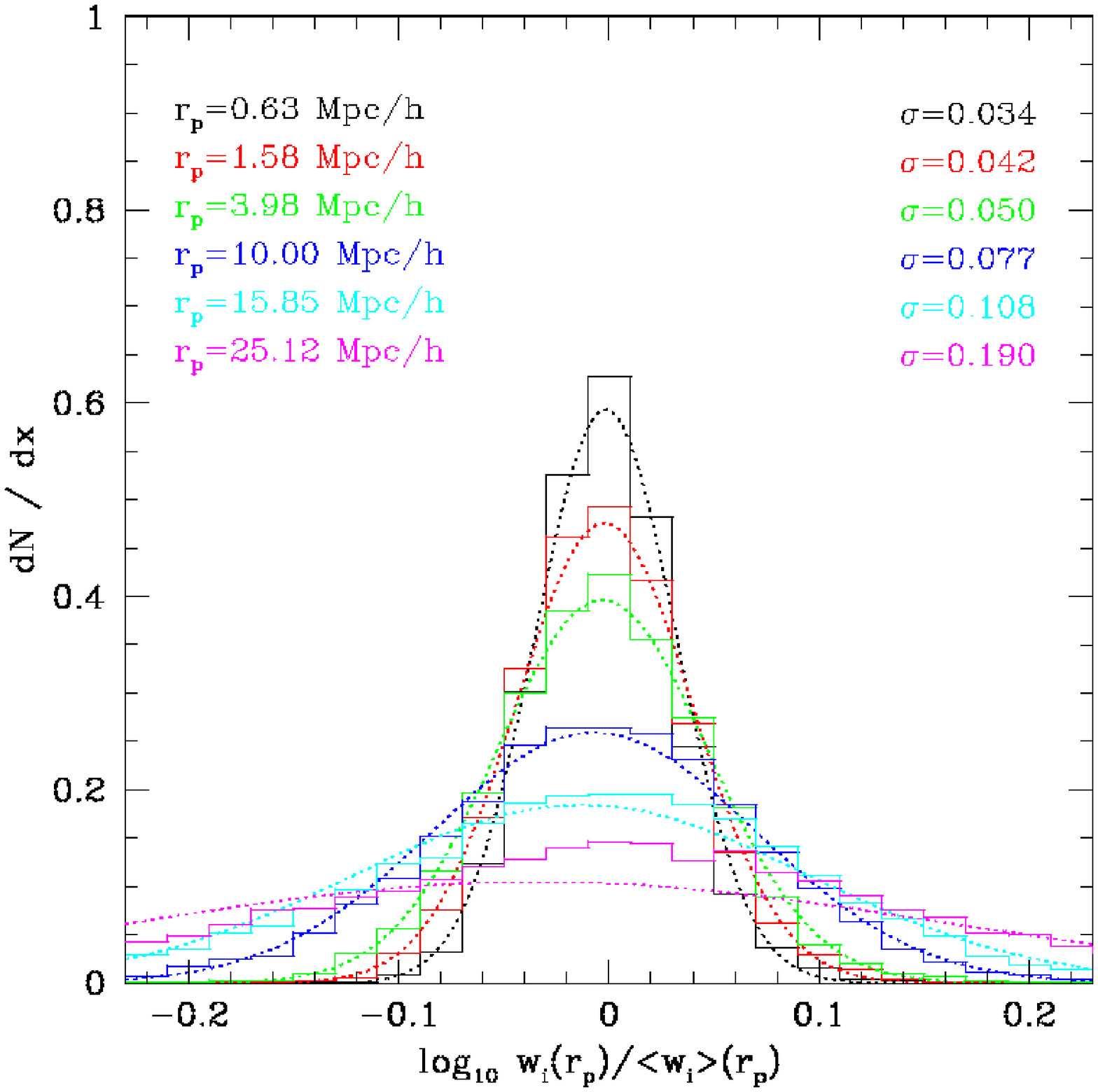}{0.33}
\caption
{
The distribution of projected correlation function measurements 
for different estimators: left: mock, centre: jackknife and 
right: bootstrap. We plot the distribution of the $\log_{10}$ of 
the measurement divided by the mean for that scale; different 
colours show the distributions for different scales, as indicated by
the legend on each panel. In the case of the mocks, the mean is the 
average over the 100 data sets. For the jackknife and bootstrap, 
the mean is averaged over the resamplings of each data set. The dotted
lines show the corresponding Gaussian distribution, with same mean and
variance, the latter indicated on the right in each panel.
In this plot, we show results for the jackknife and bootstrap 
for data sets divided up into 27 sub-volumes.
}
\label{fig:error_dist}
\end{figure*}

A closer look at the data set copies generated here with the bootstrap
method reveals that the mean effective volume of the copies is only $\sim 63\%$ of the volume of the original data set (i.e. in the case 
where the number of sub-samples selected at random with replacement 
equals the number of sub-samples the data set is divided into,
\nr=\nsub). Perhaps somewhat surprisingly, this value for the mean
effective volume depends only weakly on the number of sub-volumes into 
which the original data set is divided: in fact, for a data set split
into a large enough number of sub-volumes the mean is insensitive to
the exact \nsub\ value\footnote{Whilst the mean effective volume is 
insensitive to the number of sub-volumes the data set is divided into, 
the form of the overall distribution of relative effective volumes does 
depends on \nsub, as the only possible values are all integer fractions
of \nsub.}. 
This reduction in effective volume 
could, in principle, explain why the mean relative variance is
systematically larger for the bootstrap than that expected from the
mocks. To increase the effective volume of the bootstrap copy, we need
to sample more sub-volumes, i.e. oversample the number of sub-volumes
w.r.t. the standard case \nr=\nsub. We have experimented with
oversampling the number of sub-volumes 2, 3 and 4 times (lower half
of Table~\ref{tab:sum_runs}). As expected, the mean relative effective
volume does tend then towards unity: for \nr\ = 2~\nsub, 3~\nsub\ and
4~\nsub, the mean relative effective volumes are $\sim~87\%$,
$\sim~95\%$ and $\sim~98\%$ respectively. The effect of this on the
mean relative variance is discussed further in
\S\ref{sec:boot_resampling}. 

The shapes of the relative variance computed using the different methods 
are different as a function of scale. While $\sigma_r$ for $\xi_{r}(s)$ 
and $\xi_s(s)$ show a clear minimum between 3-6 and 2-3$\Mpc$ respectively, 
$\sigma_r$ for \wp\ shows no evidence for an increase in the relative 
variance on moving to the smallest scales considered. This dependence could 
be related to the number density of dark matter particles used in the 
analysis, with $\xi_r(s)$ and $\xi_s(s)$ being maybe more sensitive to
Poisson noise on small scales than the integral statistic \wp. The
flattening in the relative variance seen in the mocks for $\xi_{r}(s)$
corresponds to the scale where the dominant contribution to the
correlation function changes from the one-halo term (small scales) to
the two-halo term.

It is important to remember that it is not possible from
Fig.~\ref{fig:variance} alone to quantify the inevitable 
correlations between different scales in the clustering statistic. 
It would be naive to assume that these are unimportant, but let us here
for simplicity consider that naive situation of uncorrelated errors.   
In the case of jackknife errors, the different scale dependence of 
the relative variance (compared with the ``truth'') makes their
interpretation more complicated and therefore more likely to lead to
misleading conclusions, especially if the small scale clustering is
included. On the positive side, the jackknife errors, as
presented here, tend to overestimate the true errors, which is certainly
better than underestimating them. In the case of the bootstrap errors,
on the other hand, the relative variance as a function of scale has a
similar shape to that of the variance from the mocks, and so the
variance could be rescale by an appropriate factor to take into account
the overall overestimate of the relative variance. As with jackknife
errors, the bootstrap errors, as presented here, do not underestimate  
the errors on any of the scales considered, which is a good thing 
if one is to consider an uncorrelated error analysis.

\subsection{Distribution of correlation function measurements}
\label{sec:error_dist}

Having compared the estimates of the relative variance of the correlation 
function in the previous section, let us now look at the distributions 
of the measurements of the correlation functions themselves.
The motivation for doing this is clear. 
Recall that in order to be able to interpret $\chi^2$ fitting in the 
standard way, assigning the confidence intervals associated with a 
Gaussian distribution to set differences in $\chi^2$, we need to find 
a quantity which is Gaussian distributed to use in the fitting process.

The distribution of the measured projected correlation functions are 
compared in Fig.~\ref{fig:error_dist}. Each panel corresponds to a 
different error estimation method (left: mocks; centre: jackknife; 
right:bootstrap). We plot the $\log_{10}$ of the measured projected 
correlation function divided by the mean value, which shows deviations 
on either side of the mean on an equal footing. The different histograms 
show the distributions of $\log_{10}$ \wp/$<$\wp$>$ measured 
on different scales covering the range $0.60$ to $25~\Mpc$. The dotted
curves show the corresponding Gaussian distributions, i.e. with same
mean and variance.

A quick comparison of the shapes of the histograms in the different
panels shows that the distribution of $\log_{10}$~\wp/$<$\wp$>$ appear
Gaussian for all estimates on small scales, while on larger scales,
this is no longer true for jackknife and bootstrap estimates 
($r_p \gtrsim 5~\Mpc$ and $r_p \gtrsim 15~\Mpc$ respectively). 
On the largest scales considered here, even the distribution from the
mocks seems to be boarder than its corresponding Gaussian equivalent. 
The distributions yielded from the jackknife estimator tend to show
asymmetry on most scales plotted. Thus, despite these deviations from 
a Gaussian distribution, we conclude that  $\log_{10}$~\wp\ 
is closer to being Gaussian than \wp\ is, and, as such, is a 
more appropriate choice of variable to use in a $\chi^2$ fitting. 

\begin{figure*}
\plottwo{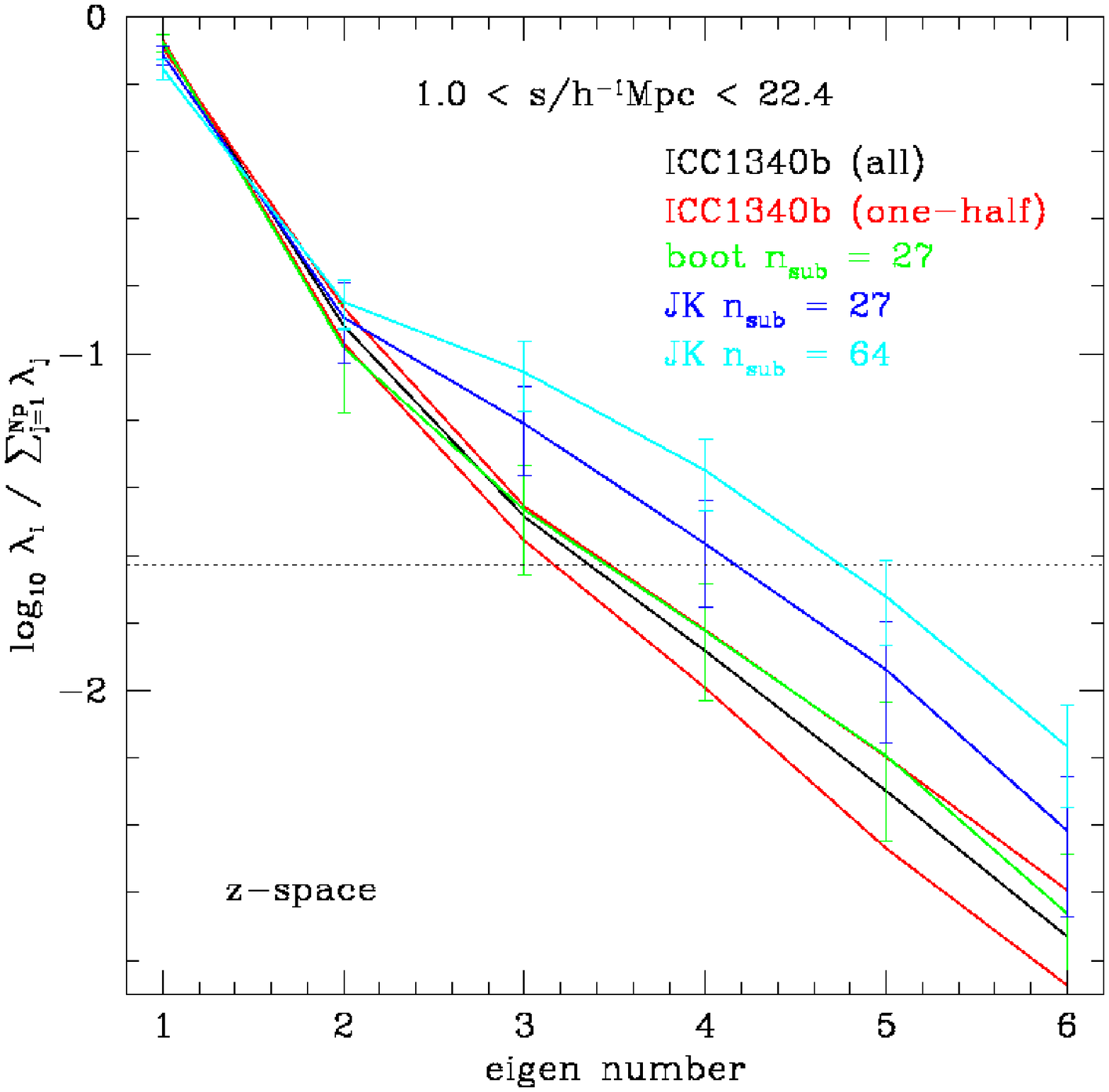}{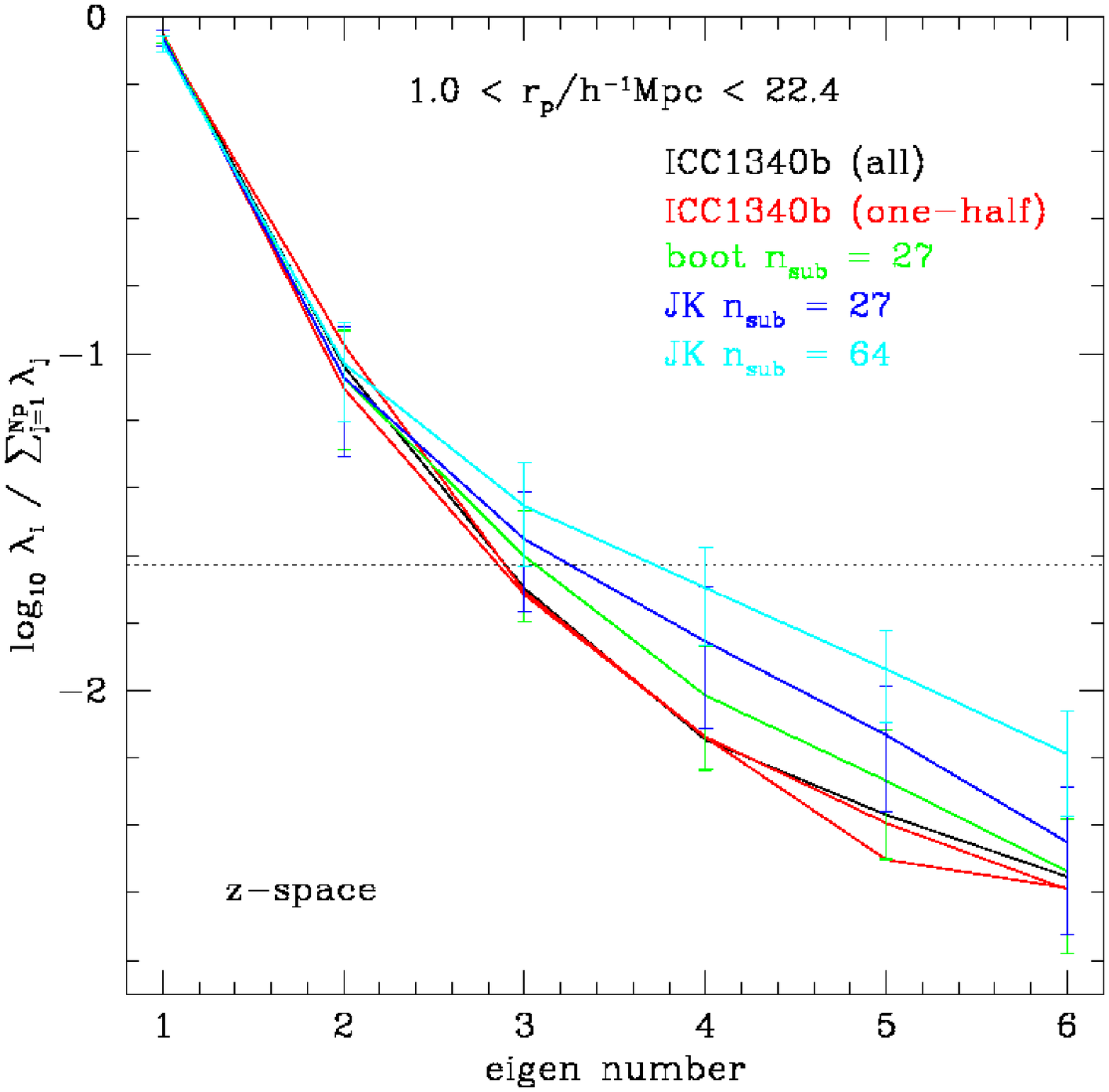}
\caption
{{\it Left :}
The normalized eigenvalues (i.e. normalized relative variance
contributions) for the redshift space correlation function, $\xi_s(s)$,
as function of eigenvalue number. The black line shows the eigenvalues
for 100 mock data sets, while the two red lines correspond to the
eigenvalues obtained from the covariance matrix obtained using the
first and last 50 mocks only. As in Fig.~\ref{fig:variance}, this
illustrates the scatter in the mock eigenvalues. The green line shows
the equivalent result for the Boot-27 errors, while the blue and cyan
lines show the Jack-27 and Jack-64 errors respectively. Errorbars
indicate the 1-$\sigma$ scatter on applying the error estimates to the
100 data sets. The range of pair separations considered in the
principal component decomposition is given in each panel. 
{\it Right:}
Same as left panel but for the projected correlation function, \wp,
estimated from the simulations put into redshift space. 
}
\label{fig:eigenval}
\end{figure*}

The middle panel of Fig.~\ref{fig:error_dist} shows the distribution of
the jackknife \wp\ estimates. 
Note that, as a result of the method of construction, jackknife 
errors are much more highly correlated than all the others, which 
is reflected by the additional factor of $N-1$ in the numerator of 
the expression for the covariance matrix (Eq.~\ref{eq:cov_jk}).
Hence, to compare the jackknife distribution with the others, we 
first need to rescale the ratio \wp\ by $\rm \sqrt{N_{sub}-1}$.
Similarly, it is essential to account for this factor when plotting 
jackknife errorbars on data points, as otherwise they do not correspond 
to the variance in the jackknife estimates. A quick comparison of the 
middle and left panels in Fig.~\ref{fig:error_dist} shows that the 
distribution of jackknife estimates over this medium range of scales is
similar to the corresponding mock distribution: the jackknife distribution is 
slightly wider than the mock on small scales, which agrees with the
comparisons of the variances presented in
\S\ref{sec:variance_rel}. However, this figure also shows a concerning
trend in that the jackknife distribution tends to become
more asymmetric on larger scales ($r_p \gtrsim 10~\Mpc$),
implying that jackknife errors for $\log_{10}$ \wp\ are not Gaussianly
distributed on such scales.

Finally, in the right panel of Fig.~\ref{fig:error_dist} we show
the distribution of bootstrap \wp\ measurements. On all scales
considered the distribution is clearly wider than that obtained from the
mocks. This highlights the fact that, when only 
considering the diagonal terms from the covariance matrix (as is 
commonly done), bootstrap errors are generally larger than both 
mock and jackknife errors, as was clear from \S\ref{sec:variance_rel}
already. However, this does not necessarily mean that bootstrap errors
overestimate the uncertainty, as we need to take into account the
correlation between errors as well. Clearly, in an analysis which
ignores the correlation of the errors, the errorbars will be
overestimated using the bootstrap method. It should be stressed that
the values of $\log_{10}$ \wp\ obtained from the bootstrap, unlike 
those from the jackknife, are certainly more Gaussianly distributed, 
with perhaps some hint of a non-Gaussian error distribution appearing 
only for $r_p \gtrsim 25$.

\subsection{Eigenvalues}
\label{sec:eigen_val}

So far we have ignored any correlations between bins when
presenting the errors. To correct for this omission, we now consider 
the first step in a principal component decomposition of the covariance 
matrix, the distribution of normalized eigenvalues. The normalized 
eigenvalues quantify the contribution to the variance of the 
corresponding eigenvector. 

In the two panels of Fig.~\ref{fig:eigenval} we present the normalized
eigenvalues for the covariance matrix of redshift space correlation 
function measurements, $\xi_s(s)$ (left), and for the projected correlation 
function, \wprp\ (right), in both cases as function of eigenvalue number. 
The eigenvalues are ranked in order of decreasing variance, with the first 
eigenvalue accounting for the largest variance. The results in this plot  
are dependent on the number of data points used. In this case, the 
correlation functions are estimated in logarithmic bins of pair 
separation with width of 0.2 dex. 

In each panel, the black line corresponds to the mock eigenvalues
(which we call the ``truth''), and the red lines indicate the
eigenvalues obtained from the covariance matrix constructed using 
just one half of all the mock realizations. This
provides an indication of the accuracy with which the mock eigenvalues
are measured. Interestingly, the accuracy of the eigenvalues
in this test is higher for \wprp\ (right) than it is for $\xi_s(s)$
(left). This is most likely related to the fact that, within the
range of scales considered, $\xi_s(s)$ is more sensitive to Poisson
noise than \wprp, as was the case for the results in Section~\ref{sec:variance_rel}.

Fig.~\ref{fig:eigenval} shows that in all cases, the eigenvalues 
decrease strongly in amplitude with increasing eigennumber. The first two 
eigenvalues alone typically account for 80 or even 90~\% of the total 
variance. This indicates that a very strong correlation exists between 
the bins in the correlation function measurements. The shape of each 
eigenvalue curve is dependent on the correlation function itself. 
Somewhat surprisingly, $\xi_s(s)$ appears to be a less correlated 
statistic than \wprp, as more eigenvalues are needed in the former 
case to represent the covariance matrix with the same level of fidelity 
i.e. to the same total variance. Or, in other words, the eigenvalue curve 
is shallower for $\xi_s(s)$ than it is for \wprp. 
Changing the range of scales or the number of data points used to 
represent the correlation function has only a marginal impact 
on the form of the eigenvalue curves.

Eigenvalues derived from the bootstrap covariance matrix are shown by 
the green line in each panel of Fig.~\ref{fig:eigenval}, with error bars 
indicating the scatter over the 100 mock data sets. Here we show only the  
bootstrap eigenvalues obtained using 27 sub-volumes.
On average, the bootstrap method recovers the expected eigenvalue curve 
rather accurately.  Although it is not a perfect match, the ``truth''
(i.e. mock) is always within the scatter presented for both clustering
statistics. Note that changing the scale or the number of data points
considered does not appear to influence these conclusions. 

The eigenvalues obtained from the jackknife covariance matrix are shown 
in Fig.~\ref{fig:eigenval} for two choices for the number of sub-volumes, 
27 and 64, shown by the blue and cyan lines respectively. 
Once again, error bars indicate the scatter in the eigenvalues 
obtained by applying this technique to each of the 100 data sets. 
Neither of the two jackknife measurements fully recovers the true 
eigenvalues. Interestingly, the more sub-samples that are used, the further
away the eigenvalue curves move from the ``truth''. For $\xi_s(s)$,
this appears to be mostly due to the much smaller first eigenvalue,
which is $\sim 0.6 \pm 0.1$ instead of  $\sim 0.85 \pm 0.1$. 
Furthermore, the slope of the eigenvalue curve is different than 
that obtained from the mock data sets and is very sensitive to the 
number of sub-volumes the data is divided into. 
We have checked that further increasing the number of sub-samples 
the data is split into simply exacerbates this problem. 
This is a first yet clear indication that the covariance matrix 
estimated from the jackknife technique is not equivalent to the 
``correct'' mock estimate. However, these issues do not necessarily
imply that the jackknife error estimate is incorrect, since the
eigenvalues are  just one part of the principal component
decomposition. On the other hand, this highlights the limitation
in how accurately the jackknife methodology can recreate the true
underlying errors, as given here by the mocks. Most of the above
remarks remain true for \wprp\ as well. As was the case with the
bootstrap analysis, changing the scale or the number of data points
does not significantly influence any of the above conclusions.

\begin{figure*}
\plottwo{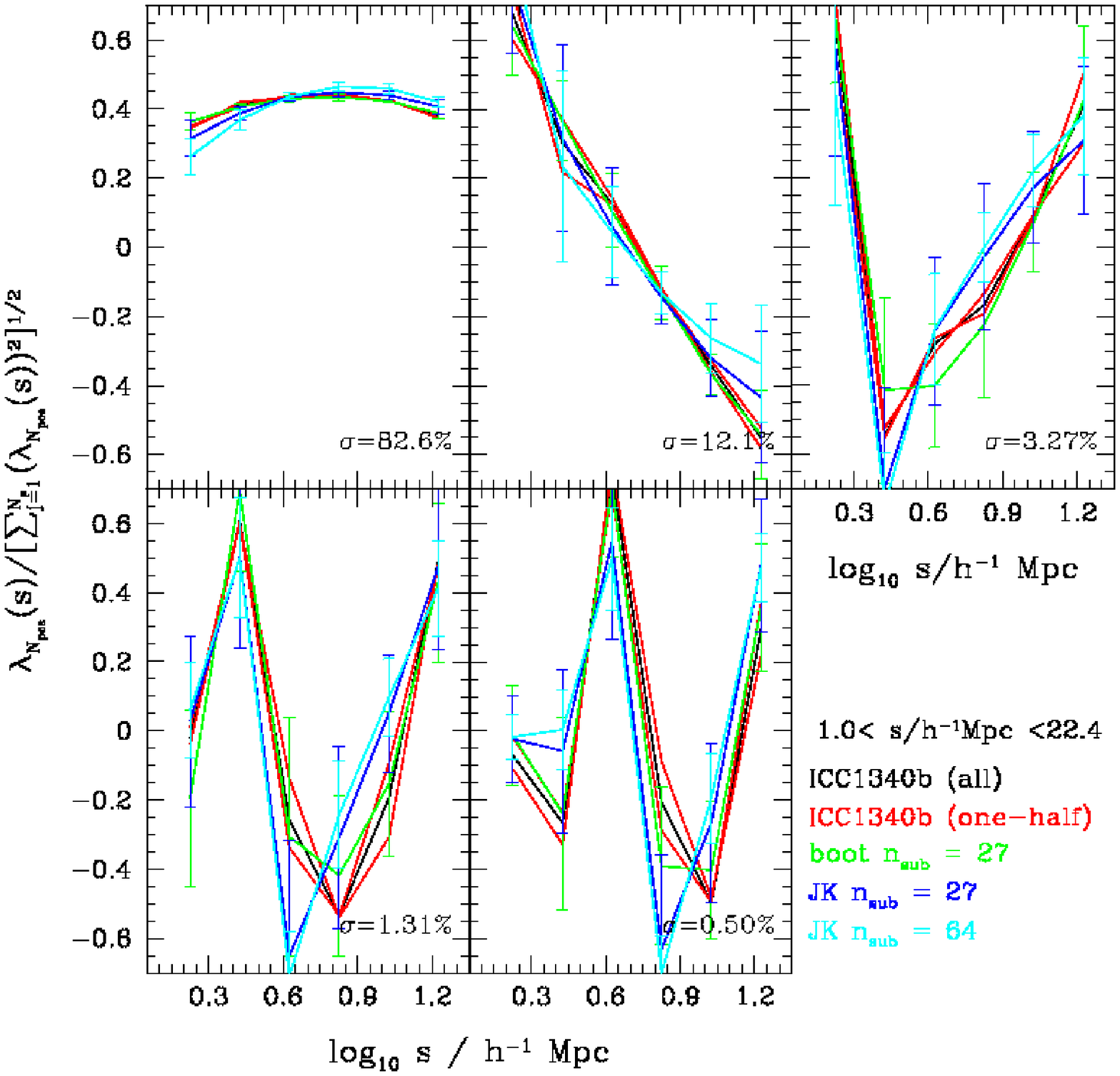}{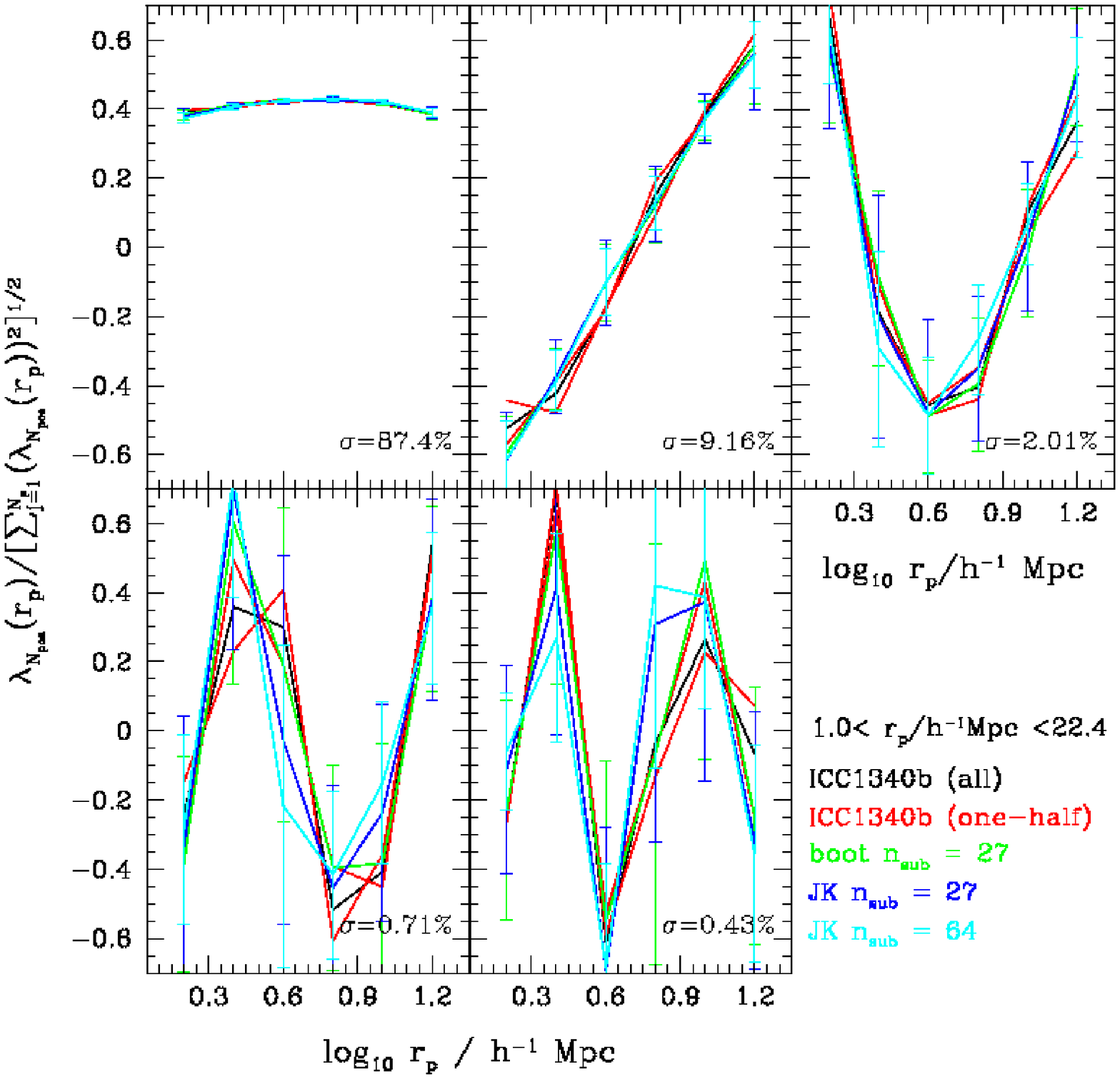}
\caption
{{\it Left:}
The normalized eigenvectors for the spherically averaged redshift-space 
correlation function, $\xi(s)$, as function of scale. Each panel 
corresponds to a different eigenvector, whose contribution to the 
total variance (in the mock case) is indicated on each panel. Colour 
and line styles are the same as in Fig.~\ref{fig:eigenval}. Error bars 
show the 1-$\sigma$ scatter obtained from the 100 data sets. The range of 
physical scales considered in the principal component decomposition 
is indicated in the legend.
{\it Right:}
Same as left panel, but for the projected correlation function, \wprp,
as function of projected separation.
}
\label{fig:eigenvec}
\end{figure*}

\subsection{Eigenvectors}
\label{sec:eigen_vec}

After considering the eigenvalues of the principal component
decomposition we now examine the associated eigenvectors. Together, 
the eigenvalues and eigenvectors completely describe the full 
normalized covariance matrix.

In Fig.~\ref{fig:eigenvec}, we show the first five eigenvectors derived 
from the covariance matrices constructed for the spherically averaged 
correlation function ($\xi_s(s)$, left), and for the projected correlation 
function (\wprp, right). Note these eigenvectors are in the same order 
as the eigenvalues presented in Fig.~\ref{fig:eigenval}. Only the
eigenvector corresponding to the smallest eigenvalue is not shown,
which contributes less than $\approx 0.3$~\% to the total variance.
The colour coding remains the same as that used in Fig.~\ref{fig:eigenval}, 
with the contribution to the relative variance indicated in the bottom 
right corner of each panel. At first sight, the results are rather
encouraging, as the mean eigenvectors for all error calculation methods
overlap reasonably well with that measured from the mocks (the ``truth'', 
shown by the black line). The scatter on the mock result is indicated 
by the spread between the red lines, which, as before, shows the results 
obtained from splitting the mock data sets into two groups of 50. 

Let us consider each eigenvector for both $\xi_s(s)$ and \wprp\ more
carefully. The first and most important eigenvector (top left corner
of both panels in Fig.~\ref{fig:eigenvec}) is very flat and is related
to the uncertainty in the mean density of objects in the data sets. 
This uncertainty causes all the clustering measurements to move up and 
down more or less coherently. To within the quoted scatter on each
clustering measurement, all of the first-ranked eigenvectors are 
identical. However, some interesting trends can already be seen in 
this decomposition. For example, increasing the number of sub-samples 
used from 8 to 27 increases the difference between the mock and
jackknife estimates, with the shape of the data inferred eigenvector
tending  to show less (more) correlation on small (larger) scales. This
is a  real effect that is further enhanced as one considers even
smaller scales than shown here. 

The second eigenvector in Fig.~\ref{fig:eigenvec}, shown in the top
middle row of both panels, displays a strong scale
dependence unlike the first eigenvector. 
The second eigenvector gives a much smaller contribution 
to the total variance, i.e. around the $\sim 10$~\% level, as opposed
to $\sim 85$~\% for the first eigenvector. The form of the second
eigenvector reveals that small scales are anti-correlated with larger
ones.
It is worth noting that all three error methods yield eigenvectors which 
look very similar for both $\xi_s(s)$ and \wprp. 
Increasing the number of sub-volumes decreases slightly the scatter in
the recovered second eigenvector, and for a fixed number of sub-samples 
the scatter is marginally smaller for the bootstrap error estimates. 
Finally, it is worth noting that the correlations as function of scale, 
despite having different slopes for $\xi_s(s)$ and
\wprp, are in fact very similar: the orientation of the 
eigenvectors is only determined up to a sign, so all the 
curves in Fig.~\ref{fig:eigenvec} can be arbitrarily 
multiplied by -1.

The remaining three eigenvectors plotted in Fig.~\ref{fig:eigenvec}
combined contribute less than a few percent of the total variance, and
the smaller their contribution to the variance, the larger is the scatter 
from the different resamplings. The fifth eigenvector (bottom right
panel) certainly tends to be dominated by point to point variations in
the data itself. This behaviour is particularly obvious when most of the
eigenvector signal appears to come from adjacent points with opposite
correlations, as seen in the lower panels for both statistics. 

Note that it is precisely this last point, whether or not the eigenvector  
provides a useful or important description of the data, that highlights 
the difficulty behind using a principal component analysis. 
When fitting a model to the data, we need to select how many 
principal components to use in the fit (see the example in 
Section~\ref{sec:test_case}). If features in the data are real 
then any theoretical model should attempt to reproduce them, and 
the eigenvectors which encode these features should be retained. 
However, if any facet of the data is dominated by noise then we 
should not attempt to reproduce it, and we should omit the 
corresponding eigenvector. One can compare covariance matrices 
without worrying about whether or not to retain different eigenvector 
components; however, this issue has a major bearing on the success of 
model fitting.  

\subsection{Stability of (inverted) covariance matrices}
\label{sec:stab_covmat}

So far we have compared the principal component decomposition of 
covariance matrices constructed using different techniques to what we 
know is the ``right'' answer - the covariance matrix obtained from a 
set of mock data sets. Unfortunately, in practice, it is rarely the 
case that one knows the correct answer beforehand. It would be useful,
therefore, 
to devise a statistic which will allow us to quantify the degree to 
which a given technique can recover a covariance matrix. Clearly no 
statistic can ever know the precision to which a measurement 
is actually correct, since that would require a priori knowledge 
of the ``truth''. But the statistic should at least be sensitive 
to the level of noise which remains in the covariance matrix estimate. 
In the absence of such a statistic at present, we consider the
usefulness of a few well-known results on the stability of errors in
our quest to determine the stability of the inversion of the covariance
matrix.

One of the easiest tests to check the stability of a covariance matrix
inversion is to repeat the whole analysis using only the odd or even 
data points (i.e. the 1st, 3rd, 5th ... data points), and to check that 
the results remain within the quoted errors, and, furthermore, 
that no systematic shift has been introduced in either case. 
Similarly, the result should remain stable to a simple rebinning
of the data.  Note that the rebinning should be done at the level of
the data itself and not on the processed results. This is to ensure
that non-linearities in the binning procedure do not introduce unwanted
biases. In the case of 2-point clustering statistics, this means that
pair counts should be rebinned. 

Porciani \& Norberg (2006) considered a quantitative test for the
bootstrap method, which consists of using $\chi^2$ values in order to
assess the stability of the recovered covariance matrix. 
Their method relies on quantifying how many components of the
principal component decomposition can be reconstructed for a fixed
number of bootstrap realizations. The figure of merit is obtained by
comparing the $\chi^2$ distributions for two different sets and numbers
of bootstrap samples, as a function of the number of principal components
used. With this method, it is possible to at least make sure that the
number of bootstrap realizations is large enough, but under
no circumstance can we tell whether the method chosen has converged to
the ``true'' covariance matrix. One nice feature of the method is that
it can actually show that certain components of the decomposition will
under no circumstance be recovered accurately enough, in which case
this gives an indication of how many components should actually be considered.

Last but not least, the results should be stable w.r.t. the number of
sub-samples considered (within moderation of course). This is something
which has to be considered in both bootstrap and jackknife analyses,
and remains probably one of the better ways to at least attempt to show
that the results obtained with internal errors estimates are as
accurate and realistic as possible.

In Section~\ref{sec:test_case}, we put the above remarks into
practice and show that each of them are able to fix or highlight
different and often problematic issues in the error analysis
using internal estimates. We note here that it is due to the
numerous mock realizations that we are actually able to discover, on a
more rapid timescale, which of the different methods is the more
promising. Certainly the task would be much more difficult in their
absence.

%
 
\begin{figure*}
\plottwo{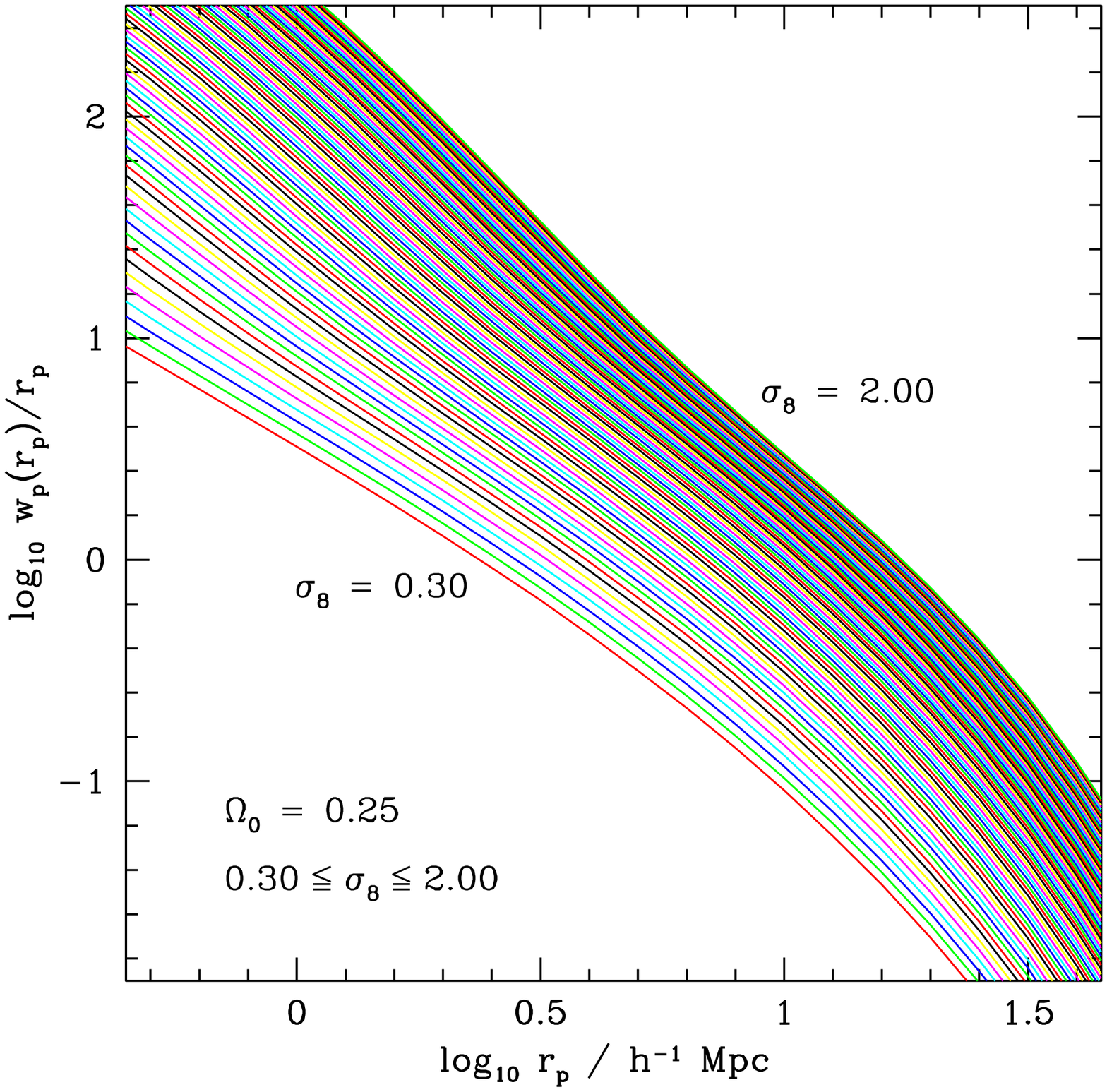}{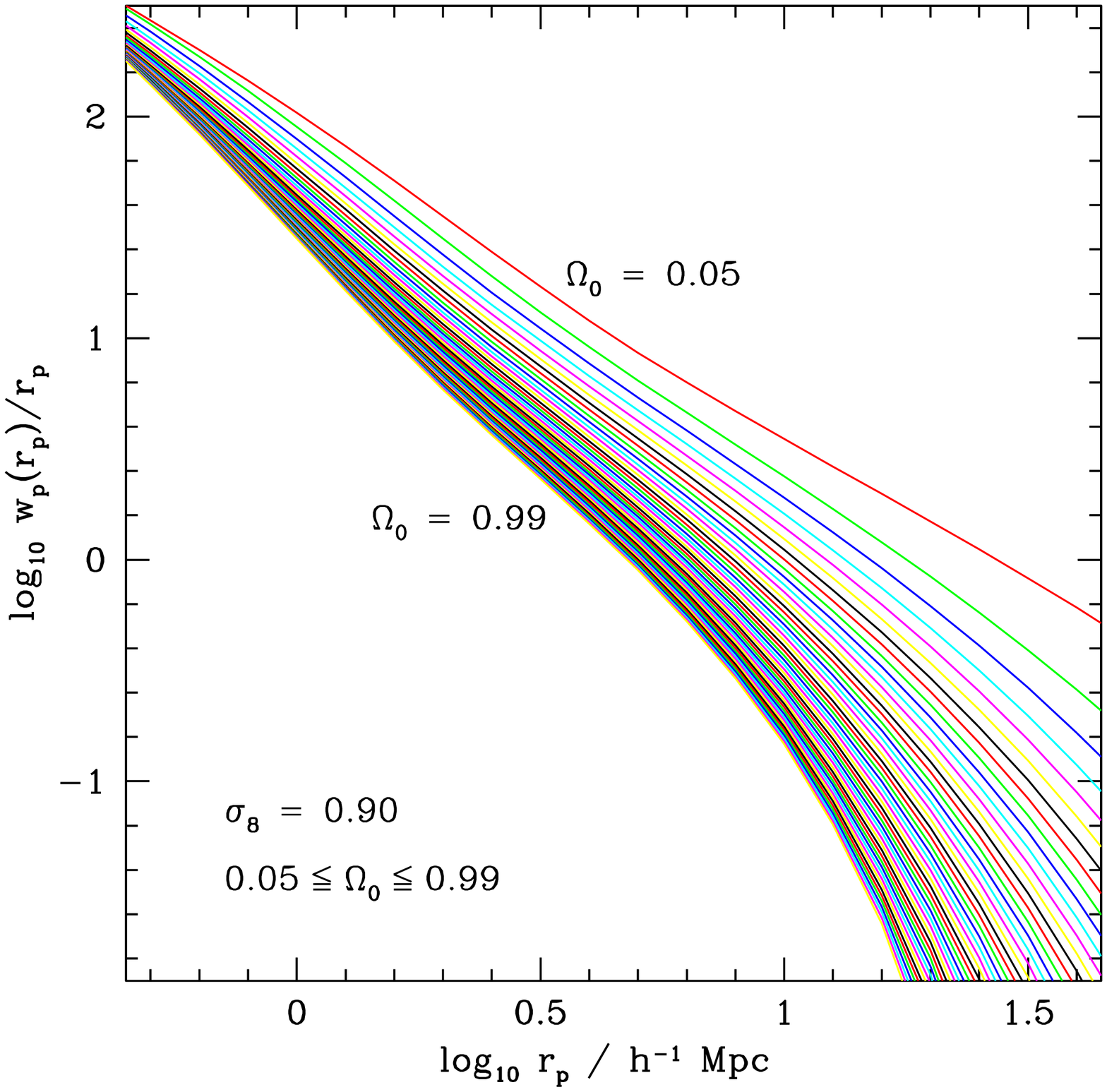}
\caption
{{\it Left:} The projected dark matter correlation function, 
estimated from a set of non-linear power spectra with $\Omega_0=0.25$ and 
$0.30 \le \sigma_8 \le 2.00$, in steps of $\delta\sigma_8 = 0.02$ (see text 
for details). Like the results presented in Section~\ref{sec:sum_runs}, the
projected correlation functions are calculated at $z=0.50$.
{\it Right:}
The same as the left panel, but now assuming $\sigma_8=0.90$ and $0.05
\le \Omega_0 \le 0.99$, in steps of $\delta\Omega_0 = 0.02$ (see text
for details).
}
\label{fig:xi_rp_model}
\end{figure*}

\section{A case study: cosmological parameter estimation}
\label{sec:test_case}

In this section we illustrate the different error estimation methods
introduced and discussed in Sections~\ref{sec:err_est}
and~\ref{sec:err_anal} with a case study involving the projected 
correlation function. The aim is to infer two basic cosmological 
parameters, the matter density, $\Omega_0$, and the amplitude of 
density fluctuations, $\sigma_8$, using measurements of the projected 
correlation function made from the mock data sets between 1 and 
$\sim 22~\Mpc$. In this example,  
because we are using data sets extracted from N-body simulations, we 
know the true values of the parameters beforehand, and can therefore 
test the performance of the error estimators. 

\subsection{A grid of theoretical \wp\ models}
\label{sec:model_grid}

We first describe how to compute a grid of models for the projected 
correlation function using linear perturbation theory. 
First, we construct a grid of linear perturbation theory power spectra, 
using the parameterization of the cold dark matter transfer function 
introduced by Eisenstein \& Hu (1998). This parameterization is an 
approximation to the more accurate results generated from Boltzmann codes 
such as {\tt CAMB} (Lewis \& Challinor 2002; see Sanchez \etal\ 2008 for 
a comparison). The initial conditions for the L-BASICC simulations were 
computed using {\tt CAMB}, so we would expect a small systematic error in 
the values of the recovered cosmological errors. However, for this 
application, the measurement errors are much larger than these 
systematic errors, and we use the Eisenstein \& Hu (1998) equations for
speed and simplicity. 

The parameters varied to construct the grid are the normalization of the 
power spectrum, $\sigma_{8}$ (over a range $=0.3$ to $2.0$ in steps of 0.02) 
and the present day matter density parameter $\Omega_{0}$ (covering 
the range $=0.05$ to $1$ in steps of 0.02). The other cosmological parameters 
are held fixed at the values used in the simulation, as described in
Section~\ref{sec:icc1340}. The power spectra are output at $z=0.5$ 
to match the redshift of the simulation output. 
The next step is to produce an approximate
estimate of the non-linear matter power spectrum, using the 
{\tt HALOFIT} code of Smith \etal\ (2003), which has been calibrated
against the results of N-body simulations. Finally, we transform the
power spectrum into the projected correlation function. The relation
between the real-space correlation function, $\xi_{r}(r)$, and the
projected correlation function is well known:
\begin{eqnarray}
\frac{w_{\rm p}(r_{\rm p})}{r_{\rm p}} & = & \frac{2}{r_{\rm p}} 
\int_{r_{\rm p}}^{\sqrt{r_{\rm p}^2+\pi_{\rm max}^2}}
\xi_{r}(r) \frac{r}{\sqrt{r^{2}-r_{\rm p}^{2}}} {\rm d} r~.
\end{eqnarray}
Using the fact that the power spectrum is the Fourier transform of the
correlation function, we can replace $\xi_{r}(r)$ to obtain, in the 
limit that $\pi_{\rm max} \rightarrow \infty$: 
\begin{eqnarray}
\frac{w_{\rm p}(r_{\rm p})}{r_{\rm p}} & = & \frac{1}{2\pi r_{\rm p}}
\int_{0}^{\infty} k P(k) J_{0}(k r_{\rm p}) {\rm d} \ln k~.
\label{eq:wrp}
\end{eqnarray}
Note that if a finite value of $\pi_{\rm max}$ is used to obtain the 
measured projected correlation function, then strictly speaking, $J_{0}$ in 
Eq.~\ref{eq:wrp} should be replaced by a finite integral over $\sin (kr)/(kr)$.
Instead we correct for the small systematic error introduced by retaining  
$\pi_{\rm max} \rightarrow \infty$ by forcing the theoretical prediction 
for $\Omega_{0}=0.25$ and $\sigma_{8}=0.9$ to agree with the measurement 
from the full simulation volume; this correction is applied to all the 
projected correlation functions on the grid and works well, which is clear 
since the returned best fitted values are still centred on the expected 
values.
The dependence of the projected two-point correlation function on 
$\Omega_{0}$ and $\sigma_8$ is illustrated in Fig.~\ref{fig:xi_rp_model}.
The left hand panel shows the dependence of \wprp\ on $\sigma_8$ for 
$\Omega_0$ fixed to the value used in the L-BASICC simulation, while the 
right hand panel shows the dependence on $\Omega_0$ with $\sigma_8$ fixed 
to the simulation value.

\begin{figure}
\plotone{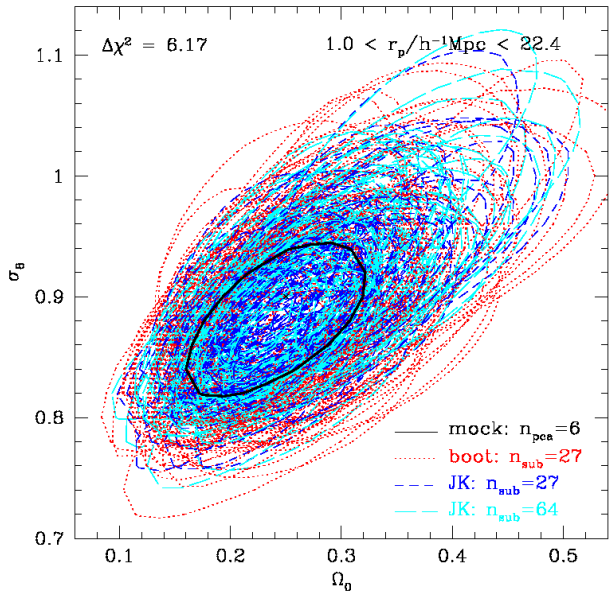}
\caption
{
$\Delta \chi^2=6.17$ contours (i.e. 95~\% confidence interval for a
  2-parameter fit in the case of Gaussian distributed errors) in the
  $\Omega_0$-$\sigma_8$ plane, for three hundred and one error
  estimates. The most realistic error estimate is given by the mock
  errors (thick black), while red (dotted) correspond to bootstrap
  errors with 27 sub-samples, blue (short dashed) to jackknife with 27
  sub-samples and cyan (long dashed) to jackknife with 64
  sub-samples. From this rather `messy' plot, it is clear that internal 
  error methods do not necessarily return the correct
  errors. The fits projection correlation function fits are done over
  the scales indicated in the legend.
}
\label{fig:om_sig_plane}
\end{figure}

\subsection{A straightforward 2-parameter model fitting?}
\label{sec:model_fit}

Armed with our grid of theoretical models we are now ready to 
test the internal error estimators against the external estimate, 
the mocks, which we regard as the ``truth''.  We note that this is 
an idealized case, which is perfect for assessing the performance 
of the internal estimators: we know the true values of the cosmological 
parameters and we know that our model should provide a very good 
description of the measurements, at least over the range of scales used 
in the fit. In a more realistic situation, the complications of 
galaxy bias, sample selection and how well we could model these effects 
would have an impact on performance (see for example Angulo et~al. 2008).  

As this is such a straightforward fitting problem, we choose in
Fig.~\ref{fig:om_sig_plane} to plot the raw 95~\% confidence interval
contours\footnote{In this section, we will refer to confidence interval
  levels as if the errors are Gaussianly distributed. Technically, we
  consider the confidence intervals (CI hereafter) which correspond 
  to $\Delta \chi^2=2.31$ or $6.17$ for 2-parameter fits, and sometimes 
  to $\Delta \chi^2=1$ or $4$ for 1-parameter fits.} 
in the $\Omega_0$-$\sigma_8$ plane for three hundred and one
error estimates. Remember that in a normal situation with 1 data set, 
we would only be able to obtain three estimates of the error, whereas in 
our case we have 100 data sets and can therefore study the distribution 
of internal error estimates.  
The black solid contour corresponds to the 95~\% confidence interval
inferred from the mocks, which we consider as the ``truth'' or the
benchmark error that the other methods are trying to match. 
The red-dotted lines correspond to Boot-27 (with \nr=\nsub), blue-short
dashed to Jack-27 and cyan-long dashed to Jack-64. The rather messy
nature of this plot tells us that there is little agreement between
each of the error methods and even between different realizations using
the same estimator.

\begin{figure*}
\plottwo{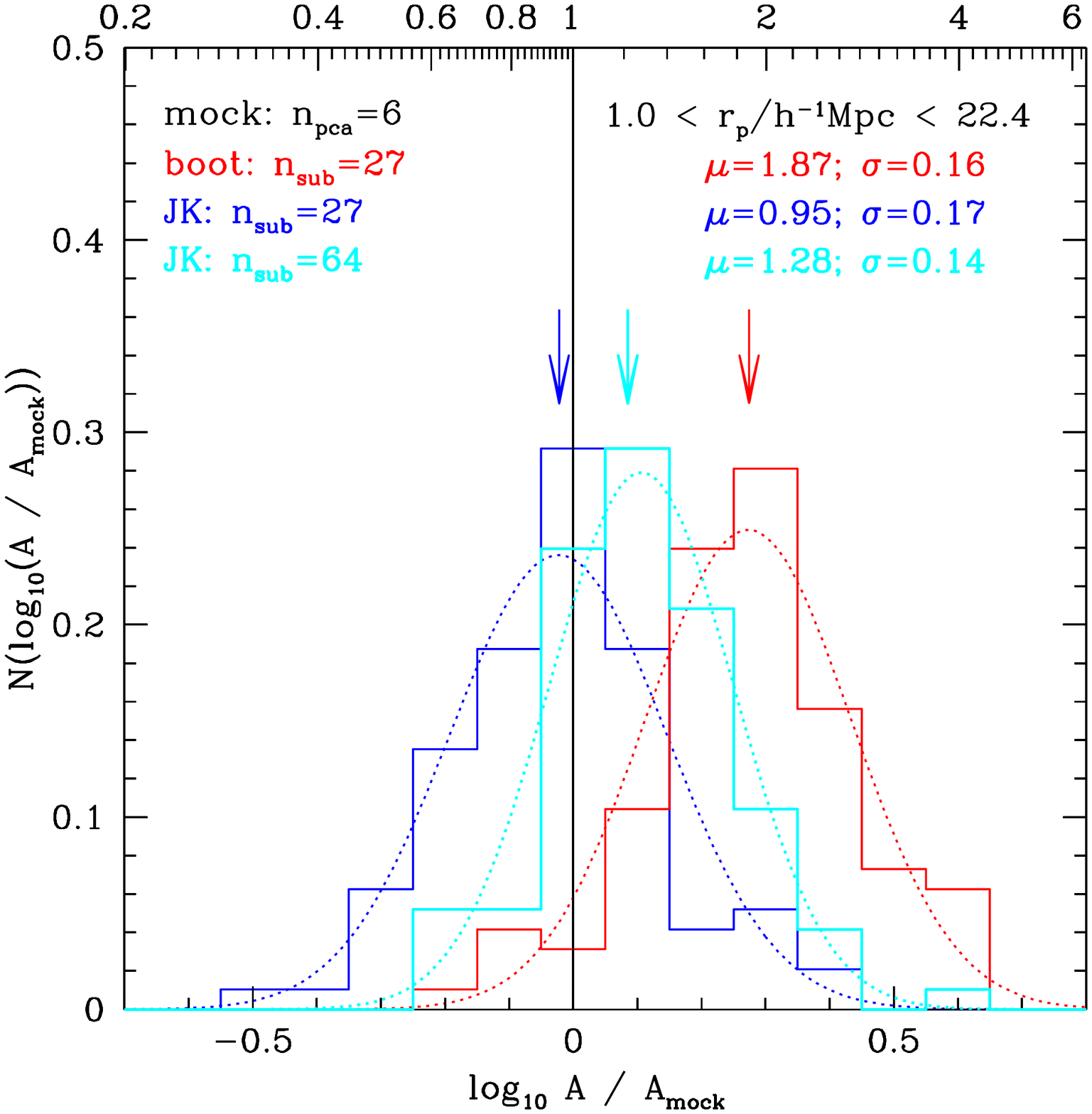}{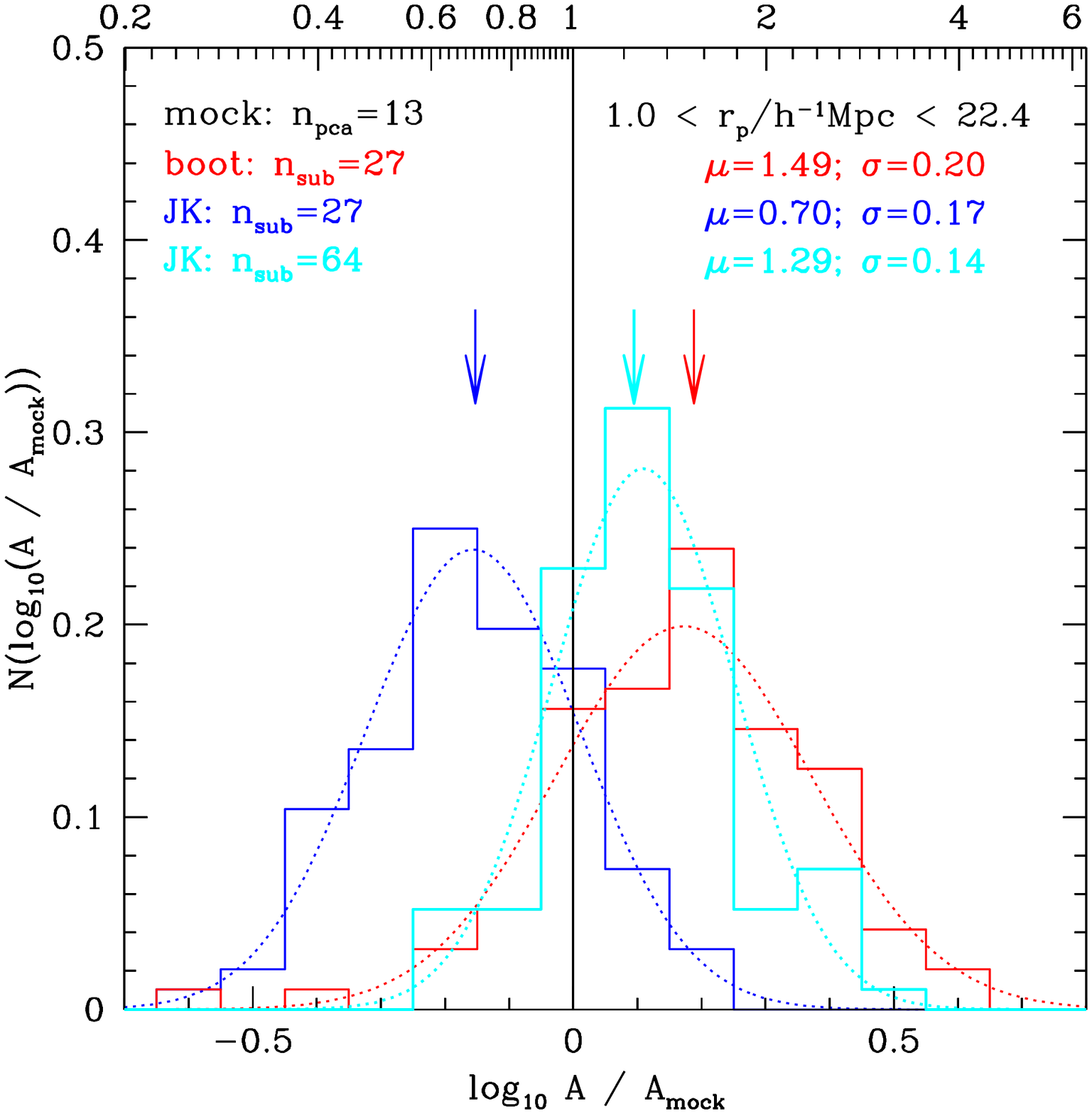}
\caption
{
The distribution of the normalized figure of merit, A~/~A$_{\rm mock}$
(see \S\ref{sec:model_fit} for details). 
Boot-27 results are shown in red, Jack-27 and Jack-64 are 
in blue and cyan respectively. The arrows indicate the median of each 
distribution, while the dotted curves show the corresponding 
Gaussian with mean and variance indicated in the panel. In the left (right)
panel, the number of principal components used is six (thirteen) out of
the six (thirteen) data points used to cover the range of scales considered. 
}
\label{fig:area_dist}
\end{figure*}

Two generic trends can be seen in Fig.~\ref{fig:om_sig_plane}. First,
not all contours are centred on the true values used in the simulations 
of $\Omega_0=0.25$ and $\sigma_8=0.90$, but appear to be systematically
shifted around. This is the effect of sample variance, also more commonly 
(but less correctly) referred to as cosmic variance. Indeed, each data set, 
despite being very large (a cube of side $380~\Mpc$), is sensitive to 
the large scale density fluctuations present in the much larger simulation 
volumes. Irrespective of the error method used, the distributions of
best fit $\Omega_0$ values are all symmetric, but not exactly Gaussian
(the central part being slightly more peaked than a Gaussian with
similar dispersion) and with mean values less than a
$\delta\Omega_0$-binsize away from the simulation value. The
distributions of best fit $\sigma_8$ values are all very well described
by Gaussian distributions, with similar dispersion and with mean values
within a $\delta\sigma_8$-binsize from the ICC1340 simulation value.
Second, some of the error contours from the internal estimators are
much larger than the mock error contour, while others are clearly much
smaller. This, on the other hand, is of concern for internal error
estimates and merits therefore closer investigation. 

To further quantify the conclusions drawn from 
Fig.~\ref{fig:om_sig_plane}, we construct a relative area
statistic, defined as the area encompassing the 95~\%
confidence interval in an internal error estimator divided by 
the area of the external, ``true'' mock error estimate, 
A / A$_{\rm mock}$, for the same number of principal components.
The motivation behind this statistic, hereafter also referred to as
the normalized figure of merit, is twofold: 1) for a 2 parameter 
model it is natural to make comparisons between different error
estimators using the full parameter plane; and 2) due to its
dimensionless nature, this statistic is easily interpreted in terms of
confidence levels (see below). A drawback of this figure of merit
is that it does not account for uncertainties in the
determination of the mock area/contours, which, given the number of data
sets available, can not be neglected. A zeroth order estimate of
that effect is given by a comparison between the following three mock
results: those obtained using all 100 mocks, the first 50 and the last
50 respectively. Additionally, it is not possible to estimate this
statistic for real data, but only for slight variants of it.

In Fig.~\ref{fig:area_dist} we plot the distribution of the relative
area statistic. The case of Boot-27 (with \nr=\nsub) is shown 
in red, results for Jack-27 and Jack-64 are shown in blue and 
cyan respectively, while the black vertical line indicates the optimal
value of the mock. The arrows show the median of each distribution. 
The difference between the two panels is the binning used for \wprp, 
and hence, indirectly, in the number of principal components
considered. 

Focusing on the left panel of Fig.~\ref{fig:area_dist} first, we see 
that for most data sets, the internal error estimates (of which three 
hundred are presented in this figure) tend to overestimate the 95~\%
CI area, on average by factors of $\sim 1.9$, $\sim 1.3$ and $\sim 0.95$ 
for Boot-27, Jack-64 and Jack-27 respectively. Furthermore, the variation  
in the area of the error contours is substantial: the central 68~\% of 
values is typically spread over 0.15 in log$_{10}$~A~/~A$_{\rm mock}$, 
i.e. a factor of $\sim 1.4$ in the area of the error contour. 
Hence, even for this particularly simple case, the uncertainty on the 
internal estimate of the CI is large and certainly not negligible. 
As seen in earlier cases, the difference between Jack-27 and Jack-64 
is also quite large, with Jack-64 yielding a marginally more centrally 
concentrated distribution than Jack-27, whereas Jack-64, on average, 
overestimates the true area by an additional 30~\%. The situation is 
clearly worst for bootstrap errors, which display by far the largest 
systematic offset, but with a spread that looks amazingly similar to 
jackknife results. Interpreting these offsets in the framework of
Gaussian errors for this figure of merit, we conclude that on
average a 95~\% boot-27 CI corresponds effectively to a 99.6~\% CI, a
95~\% Jack-64 to a 98.0~\% CI, and a 95~\% Jack-27 CI to a $\sim$90~\%
CI. This is without taking into account the spread in relative areas
from different realizations.

Unfortunately, this is not the end of the confidence interval story. We
now consider the right hand panel of Fig.~\ref{fig:area_dist}, which is
equivalent to the left except that 13 bins instead of 6 have been used
for \wprp\ (over the same range of scales) and hence all 13 principal
components are used in the fitting, instead of just 6 as in the left
panel. The difference in the figure of merit between right and 
left panels is clear, with the offsets of two of the three error 
methods changing in a systematic way. The right panel area ratios are 
biased by factors $\sim 1.5$, $\sim 1.3$ and $\sim 0.7$ for Boot-27, 
Jack-64 and Jack-27 respectively. The only positive note is that all 
distributions plotted are well described by Gaussians, shown by the 
dotted lines, with small dispersions. Indirectly, we conclude that on
average an estimated 95~\% CI corresponds, with 95~\% confidence, to a
true CI in the range 79.2~\%~-~99.8~\% for the case considered
here. We note that only Jack-64 seems to remain stable w.r.t. the
change in binning. This is most likely related to the results found by
Hartlap \etal\ (2007). They showed that internal error estimators
tend to be systematically biased low on average when the ratio of the
number of bins considered to the number of samples becomes too large
(typically 20 \% or more).

\begin{figure*}
\plottwo{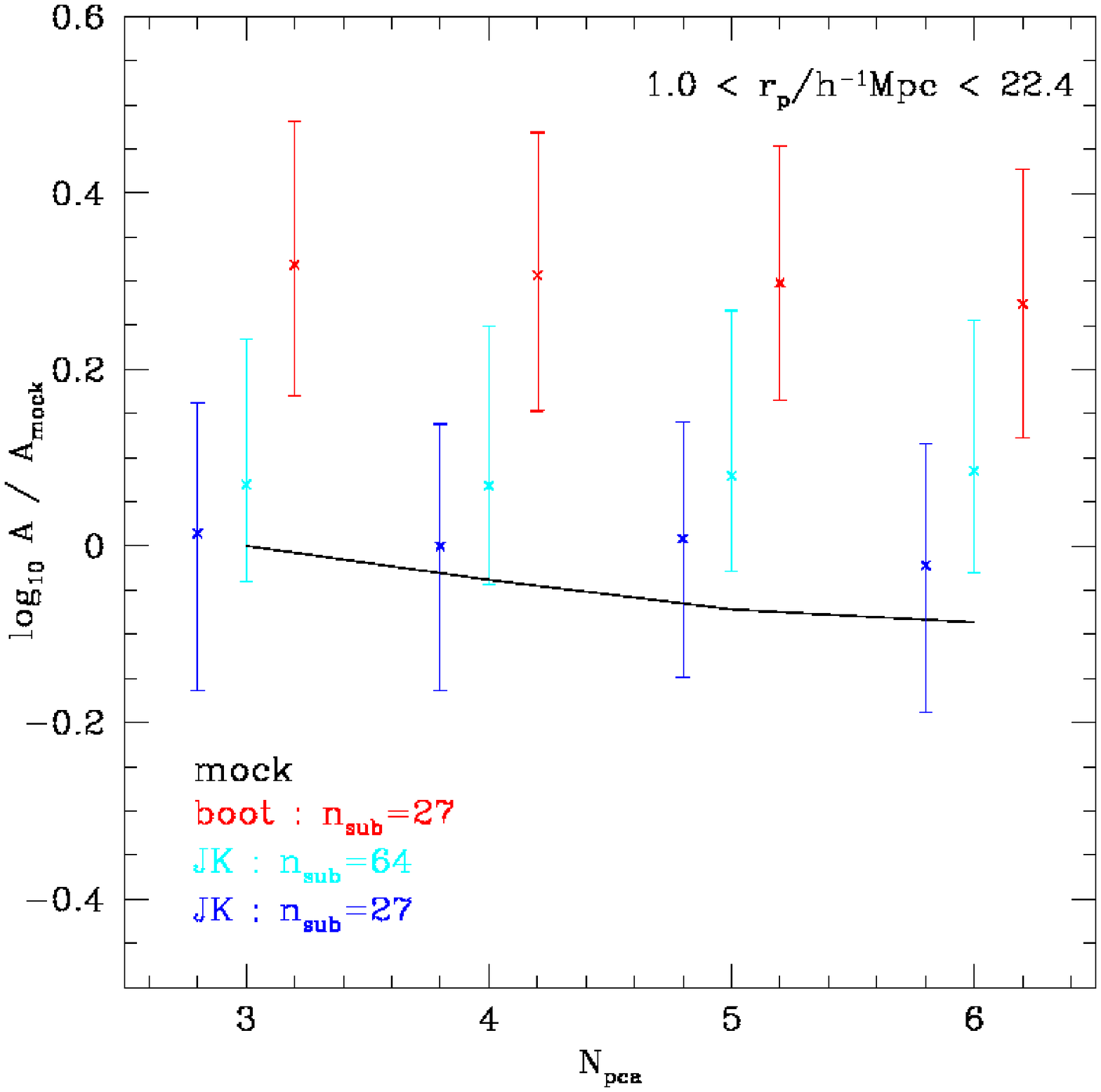}{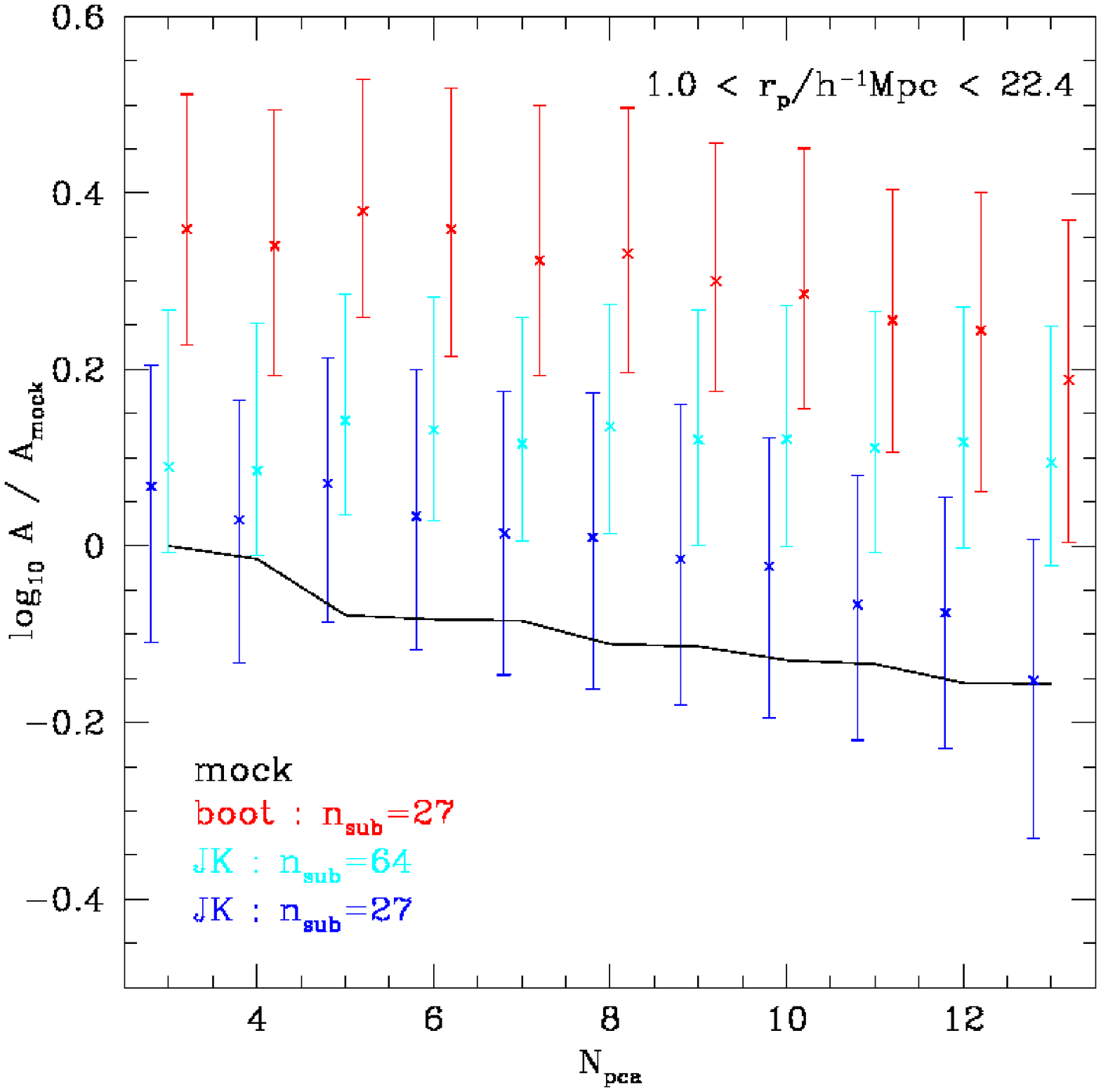}
\caption
{
The 16$^{\rm th}$, 50$^{\rm th}$ and 84$^{\rm th}$ percentiles of the
relative area statistic (see \S\ref{sec:model_fit} for a definition) 
as function of the number of principal components used in the fit: 
in each case the figure of merit is normalized to the mock area with 
the same number of principal components. 
Bootstrap errors from 27 sub-volumes (with \nr=\nsub) are shown 
in red, while jackknife errors are given in blue and cyan for 27 and 64 
sub-volumes respectively. In the left (right) panel, six (thirteen) 
equally spaced data points for \wprp\ are considered in the model
fitting and cover the full range of scales indicated in the
key. Finally, the black line indicates 
$\rm A_{\rm mock}^{N_{\rm pca}} / A_{\rm mock}^{N_{\rm ref}}$ with 
N$_{\rm ref}=3$ in both panels. 
}
\label{fig:area_dist_npca}
\end{figure*}

The differences between the distributions of error contour areas shown 
in the left and right panels of Fig.~\ref{fig:area_dist} is  
disconcerting, since only the number of data points used has changed 
(i.e. the binning of \wprp\ over a fixed range of scales). For mock
errors, we measure an increase of 15~\% in the area covered by the
95~\% confidence intervals when changing from 6 to 13 data points.
This is much smaller than the systematic shifts displayed by the
internal error estimates under the same circumstances, which present a
30~\% to 80~\% change. It is unlikely that these strong shifts are due
to the nature of the data alone, but much more likely to be due to
errors propagating through the full covariance matrix analysis. In both
panels, the fits are performed using the full covariance matrices, and
therefore we investigate in the next section if noise in the covariance
matrix is the root of the problem.

\subsection{Dependence of confidence intervals on N$_{\rm pca}$} 
\label{sec:model_fit_npca}

Perhaps the large variation in the error contours returned by the
internal estimators is due to noise in the covariance matrix, resulting
from too many principal components being retained in the analysis.

To test this idea, we show in Fig.~\ref{fig:area_dist_npca} the
impact on the relative area statistic of varying the number of principal 
components for a fixed number of data points. In the left panel of
Fig.~\ref{fig:area_dist_npca}, 6 data points are used 
to represent \wprp. The variation of the 95~\% confidence interval area
in the external mock estimate w.r.t. the area with N$_{\rm pca}=3$ is
plotted as a black line. This is a measure of how stable the mock results
are as function of the number of principal components. There is a modest 
reduction in the size of the error contour, $\sim 0.1$dex, on using 
6 principal components instead of 3, as shown by the shallow slope 
of the line. The situation is slightly more unstable in the right
panel (where 13 data points are used instead of 6 as in the left panel), 
with up to a 40~\% change over the full range of numbers 
of principal components considered. This variation is significantly 
reduced if one starts with at least 5 principal components, where 
the change in the figure of merit over the full range is then 
less than $\sim 20$~\%. 

In Fig.~\ref{fig:area_dist_npca} we also show the median, and 
16$^{\rm th}$ and 84$^{\rm th}$ percentiles of the relative area
distributions for Boot-27 (with \nr=\nsub), Jack-27 and Jack-64
estimates in red, blue and cyan respectively. The impression gained 
from the two panels is rather different, especially for Boot-27 and 
Jack-27. While the main properties of the relative area distributions
for those two estimators remain roughly the same as function of 
N$_{\rm pca}$ in the left panel (very weak dependence on 
N$_{\rm pca}$), they change systematically and significantly with
N$_{\rm pca}$ in the right panel. These changes result in unstable
error estimates, most likely due to the propagation of a bias through
the covariance matrix estimate (see Hartlap \etal\ 2007). Only Jack-64
seems to return a somewhat stable results as function of N$_{\rm pca}$.

\begin{figure*}
\plottwo{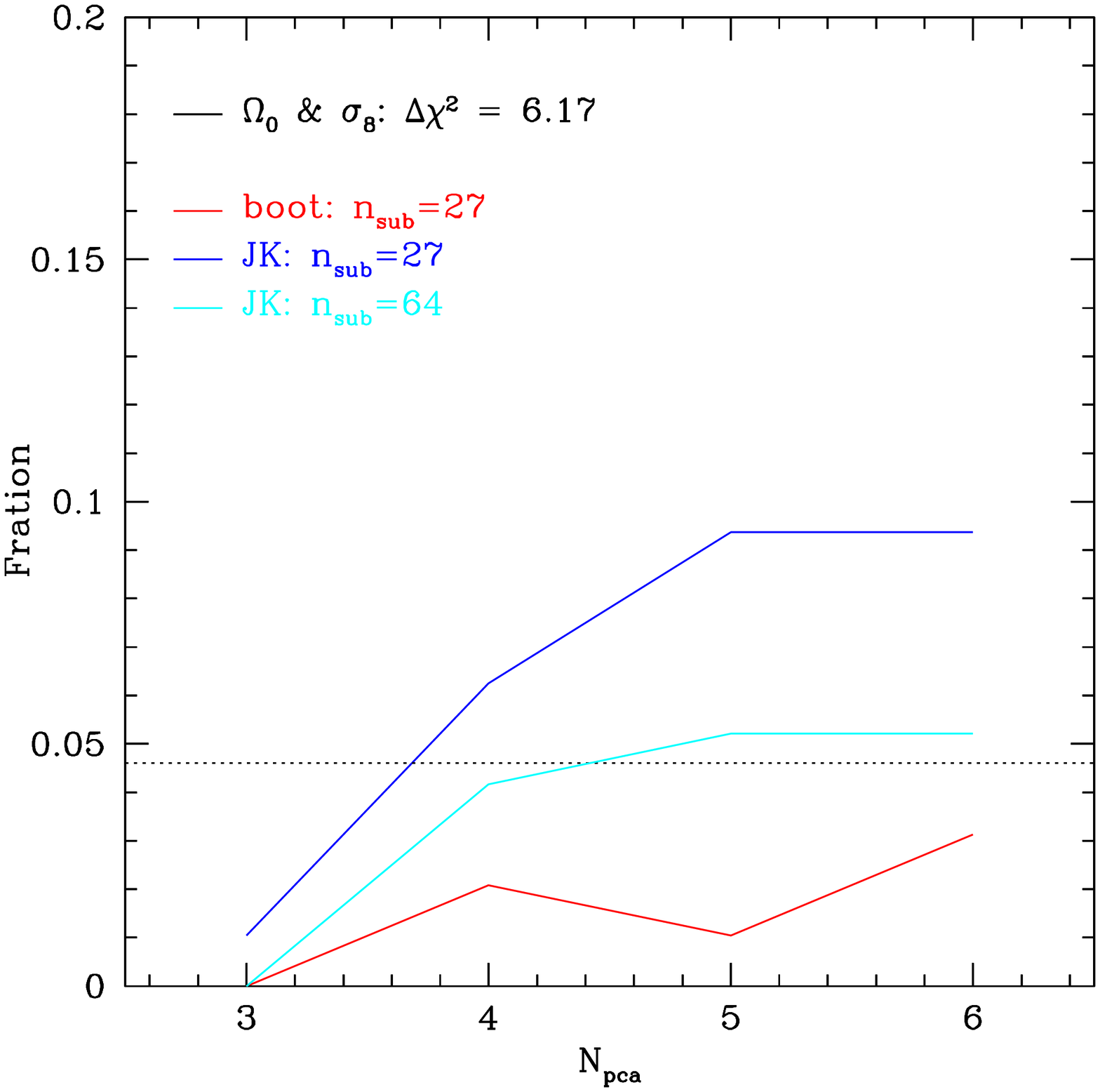}{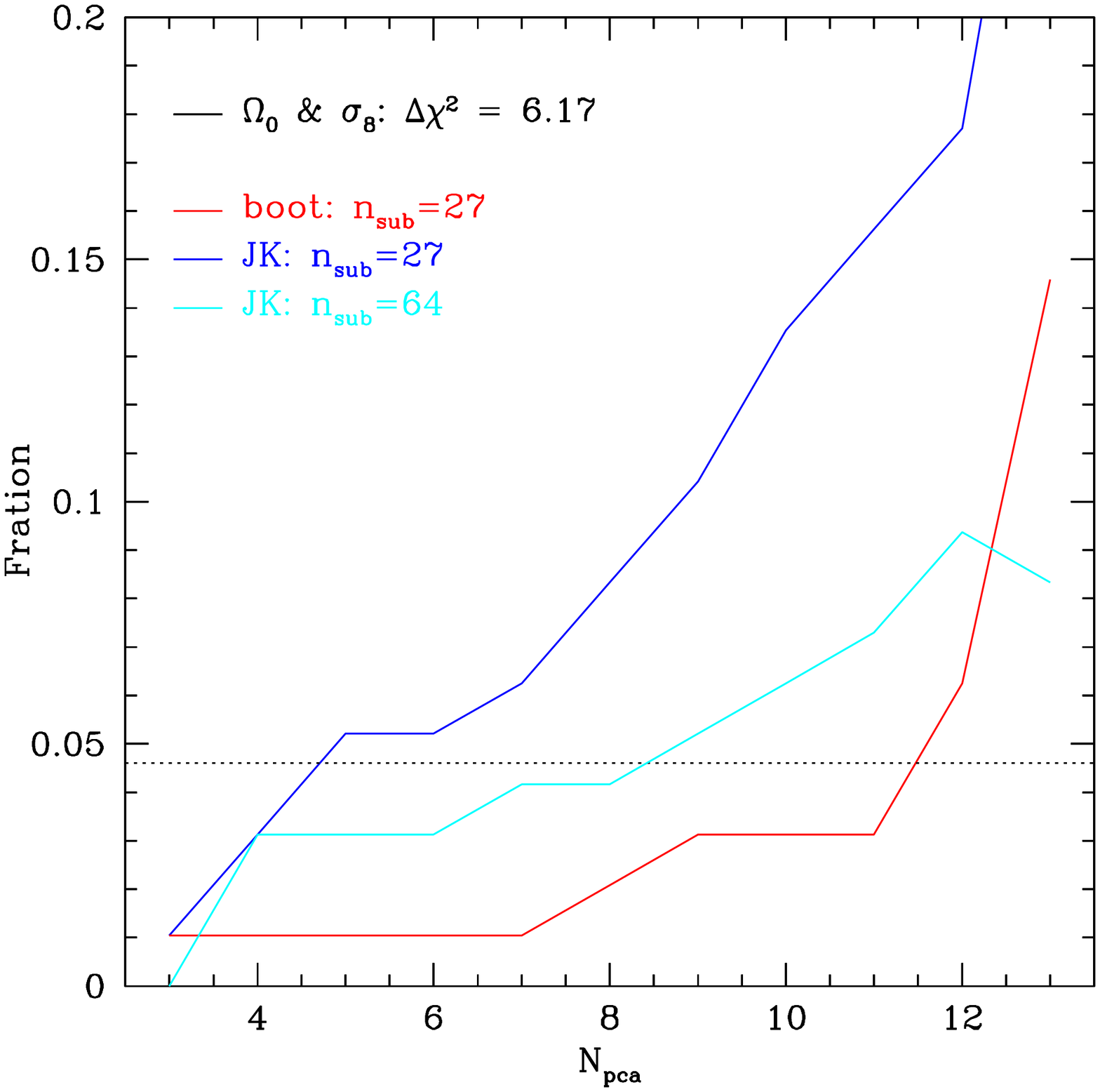}
\caption
{
The fraction of outliers as function of the number of principal
components used in the fit. An outlier is defined as a 2-parameter fit
for which the $\Delta \chi^2=6.17$ contour of the data inferred
errorbar does not enclose the true underlying cosmology,
i.e. $\Omega_0=0.25$ and $\sigma_8=0.90$ (see \S\ref{sec:icc1340}). The
horizontal line corresponds to the theoretical expectation for the
contour $\Delta \chi^2=6.17$, i.e. 95~\% assuming Gaussianly
distributed errors. The left (right) panel corresponds to bins of 0.2
(0.1) dex width. 
}
\label{fig:dist_fit_para}
\end{figure*}

Finally, in Fig.~\ref{fig:dist_fit_para} we examine the fraction of
``outlier'' experiments as function of the number of principal
components used in the fit. An outlier is defined as a data set for
which the error contour from an internal estimate (defined by 
$\Delta \chi^2=6.17$) does not include the true underlying cosmology
(i.e. $\Omega_0=0.25$ and $\sigma_8=0.90$). For a Gaussian
distribution, we would expect no more than 5 ``outliers'' defined in
this way out of 100 data sets. The left panel looks reasonable for most
N$_{\rm pca}$, especially considering the number of realizations
available: remember we have just 100 data sets at our disposal, so the
Poisson noise on the expectation value of this statistic is
non-negligible. There 
is still a tendency from this panel to say that Boot-27 (with
\nr=\nsub) overestimates the errors and Jack-27 underestimates
them. This is in relatively good agreement with our conclusions from
Fig.~\ref{fig:area_dist_npca}. In the right hand panel the situation is
significantly different. First all three estimators present radical
changes in the number of ``outliers'' as function of 
N$_{\rm pca}$. Clearly we have to exclude fits which use too many
principal components: a cut can therefore be made at about 10 principal
components for Boot-27 and Jack-64, while the cut has to be made at
about 6 or 7 components for Jack-27 already.

Neither Fig.~\ref{fig:area_dist_npca} nor Fig.~\ref{fig:dist_fit_para}
can be made without the full battery of data sets we have, nor without
the knowledge of what the ``truth'' is. However, we can nevertheless 
learn some simple tricks which will be useful for the analysis of real 
data. For example, for each internal error estimator we can plot the
relative area statistic, by using a fiducial reference value for the
area. If this quantity varies in a significant manner as function of
N$_{\rm pca}$ or as a function of the internal error estimator used, we are
made aware of a change of regime, without necessarily knowing what to
do. The most likely and robust solution will be to introduce less
components in the fit.

\subsection{Important considerations for error analyses}
\label{sec:summary_test_case}

We summarize here a few results from our simple case study. To recap the 
exercise consisted of fitting the projected correlation function (on scales
between 1 and $\sim 22~\Mpc$), using what we knew beforehand was 
the {\it correct theoretical model} and which was only dependent on 
the parameters $\Omega_0$ and $\sigma_8$, to one hundred
projected correlation function estimates, taken from one hundred data
sets of $380~\Mpc$ on a side, extracted from 50 totally independent
N-body simulations of $1340~\Mpc$ on a side. The main conclusions for
internal error estimates are, with confidence interval abbreviated as CI:
\begin{itemize}
\item the average systematic offset encountered on an estimated
  2-$\sigma$ CI implies it can be mistaken for a 1.6-$\sigma$ to
  2.9-$\sigma$ CI in reality.
\item the spread in errors for all internal error estimates is
  large: for an unbiased estimator, a 2-$\sigma$ CI corresponds to
  a real CI in the range 1.3-$\sigma$ to 3.1-$\sigma$, at the 95~\% 
  confidence level.
\item the CI from bootstrap errors, estimated with \nr=\nsub, tend to be
  systematically more biased than jackknife errors.
\item the true 95~\% CI, as measured from one hundred mocks, is
  known to $\sim 20$~\% accuracy. 
\end{itemize}
The above remarks depend on the mean number density of the sample, here
fixed to mimic a \lstar\ galaxy sample (see
\S\ref{sec:icc1340}). In the next section, we indirectly show how the
amplitude of the errors scales w.r.t. the mean number density
considered.

\begin{figure*}
\plottwo{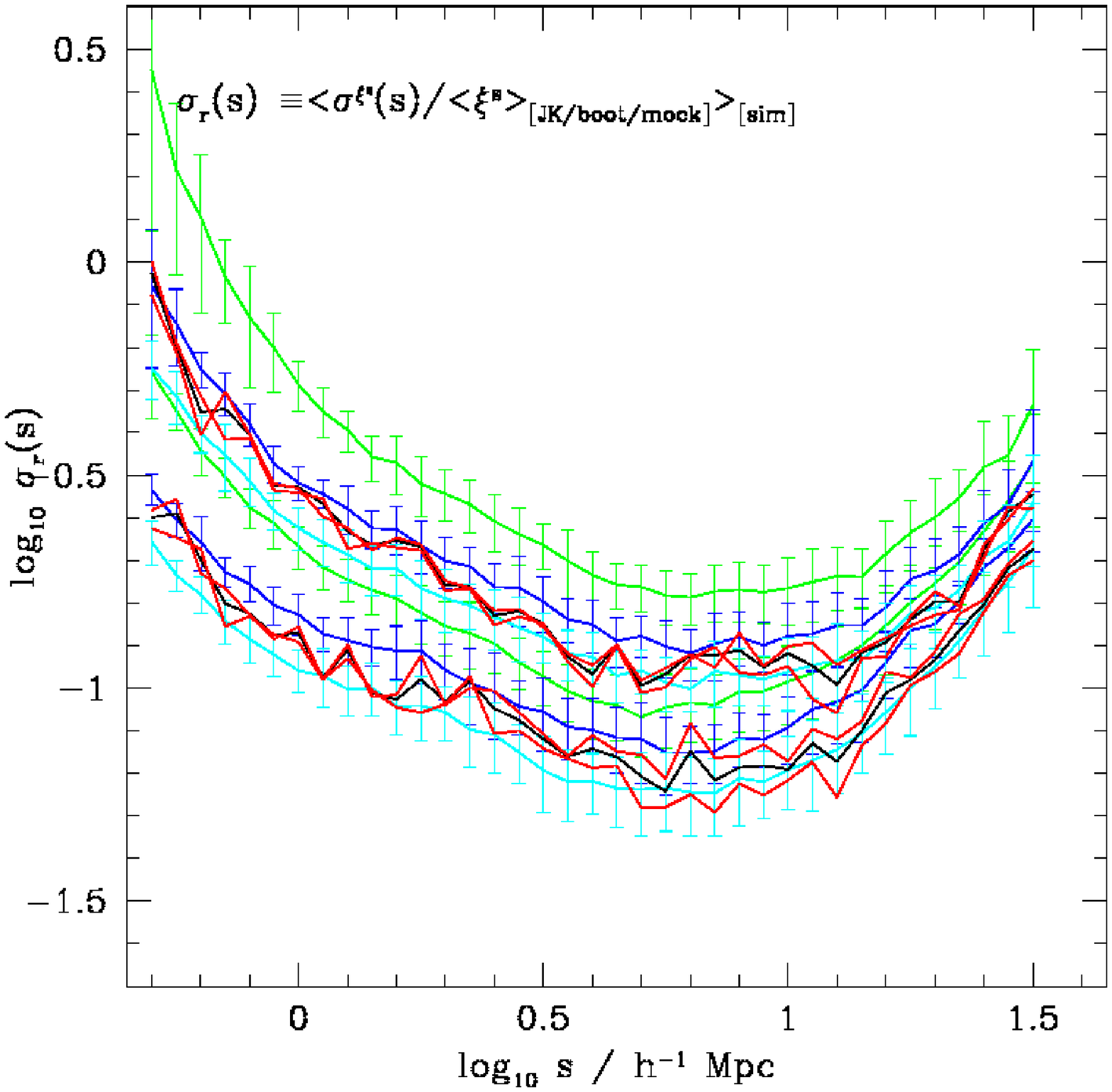}{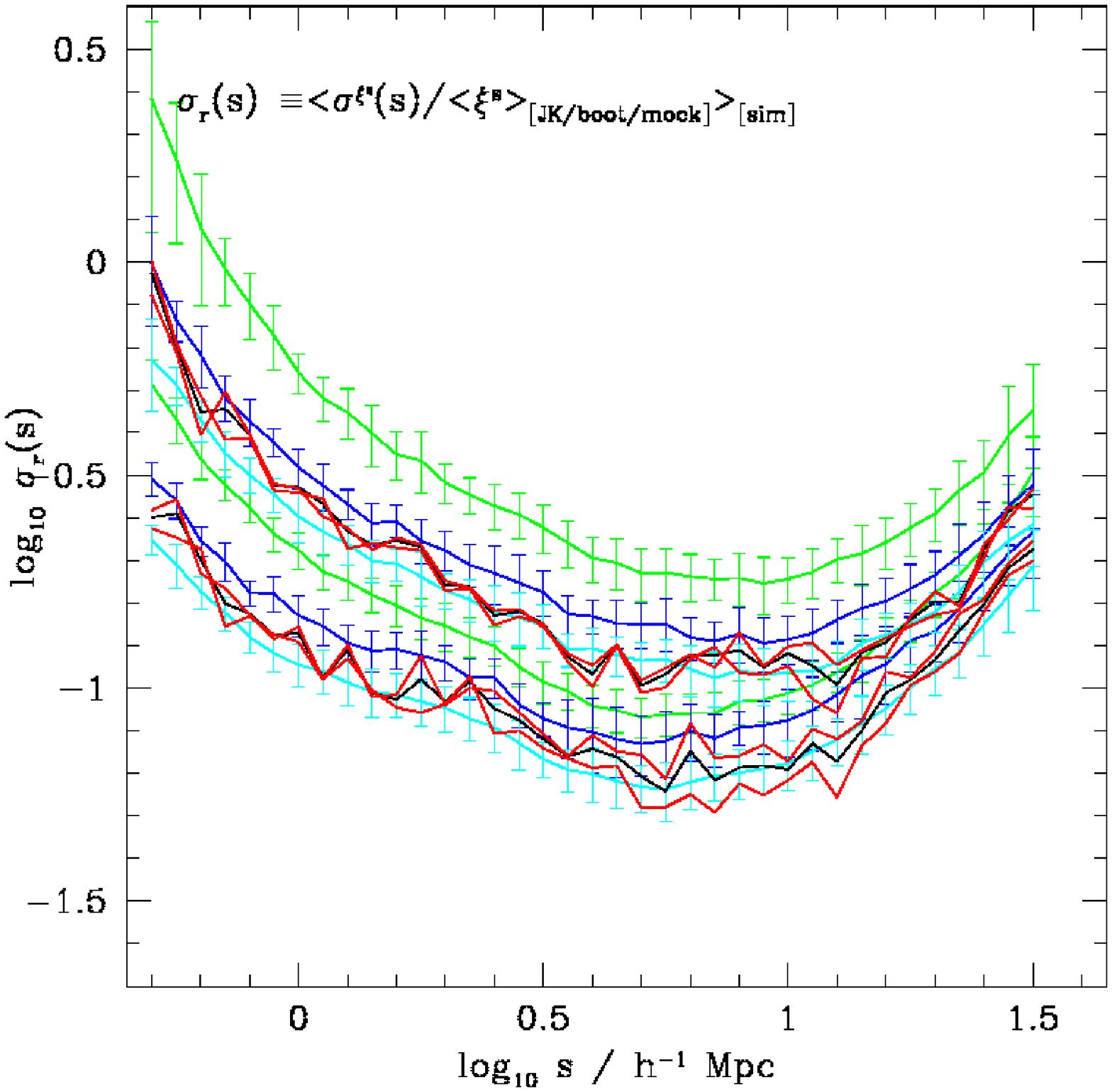}
\caption
{
Relative variance for $\xi_r(s)$ (to be compared with the left panel of 
Fig.~\ref{fig:variance}). {\it Left:} for \nsub$=27$, we compare, for two 
sampling fractions (upper group of lines: 10~\%, and lower group of
lines: and 25~\%), the  
relative variance for different number of bootstrap 
resamplings: \nr$=1$, 2, and 3~\nsub\ from top 
to bottom (green, blue and cyan lines respectively). The 
black and red lines have the same meaning as in 
Fig.~\ref{fig:variance}, with the top (bottom) ones 
corresponding to a sampling fraction of 10~\% (25~\%).
{\it Right:} same as left panel, but for
\nsub$=125$. 
}
\label{fig:boot_resampling}
\end{figure*}

\subsection{An improvement to the bootstrap recipe}
\label{sec:boot_resampling}

Throughout our comparisons we have seen that the bootstrap  
gives errors that are systematically larger than the ``truth'' 
(see for example Figs.~\ref{fig:variance} and~\ref{fig:area_dist}). 
Previously, we have set \nr=\nsub, i.e. the number of sub-volumes 
chosen with replacement was equal to the number of sub-volumes each
data set is divided up into. This is an arbitrary choice and there is no 
reason why we cannot increase the number of sub-volumes used to define 
each realization of the data set. Here we consider varying the number 
of sub-volumes chosen. We also consider increasing the number of
sub-volumes the data set is split into, which affects both the
bootstrap and jackknife methods (recall that in the jackknife, the
choice of \nsub\ sets the number of realizations we can generate).

In the left panel of Fig.~\ref{fig:boot_resampling} we show the impact
of changing the number of sub-volumes chosen on the relative variance
obtained from the bootstrap with \nsub$=27$. Two families of 5
different coloured curves are shown: 
the top ones correspond to a sampling fraction of 10~\% of the standard
mean number density, while the bottom ones corresponds to a 25~\%
sampling fraction. Each family is composed of three bootstrap samples
(\nr=1, 2, 3~\nsub, in green, blue and cyan respectively) and three
mock curves: all mocks in black, and two red curves for the 50 first
and 50 last respectively. The relative variance decreases as 
the number of sub-volumes selected increases, with the biggest change
coming when the number of sub-volumes is oversampled by a factor of
two (blue line). With an oversampling of a factor of 3 (cyan line), the
bootstrap errors are in very good agreement with those derived from the
external estimate. If we increase the oversampling rate further, the
relative variance returned by the bootstrap becomes too small (not
shown for clarity). From this figure, an oversampling factor of about
three seems to be optimal, i.e. selecting at random, with replacement,
\nr=3~\nsub\ sub-volumes from the original list of length \nsub.

We note that the left panels of Figs.~\ref{fig:variance}
and~\ref{fig:boot_resampling}, in which the density of objects in the
data set is varied, give extremely  
valuable information on how the relative errors scale for samples with 
different mean densities, as together they span an order of magnitude
variation in the mean density. Increasing the mean density by a factor
2.5 leads to a reduction in the relative variance error by $\sim 80$~\%
(top versus 
bottom black lines in Fig.~\ref{fig:boot_resampling}), while a mean
density 4 times larger decreases the relative error by a further 50~\%, 
at which stage the change becomes more scale dependent. Indeed, the
shape of the mean relative variance in the left panel of 
Fig.~\ref{fig:variance} is slightly different to those in
Fig.~\ref{fig:boot_resampling}. We attribute this to the fact that
there is a limiting number density beyond which the errors
are not any longer dominated by sample shot-noise.

The right panel of Fig.~\ref{fig:boot_resampling} shows the same
information as the left panel, but for a different number of 
sub-samples: 125 instead of 27. The remarkable point to realize here is
that both panels are virtually identical, leading 
to the conclusion that the precise number of sub-samples
the data is split into is not a primary factor for determining 
the size of bootstrap errors. 
This is in strong contrast to jackknife errors for which we observed, 
in Fig.~\ref{fig:variance}, a systematic change in the relative variance 
as function of scale with respect to the mocks, with increasingly
discrepant results on smaller scales for the jackknife analysis with
increasing number of sub-samples.

\begin{figure}
\plotone{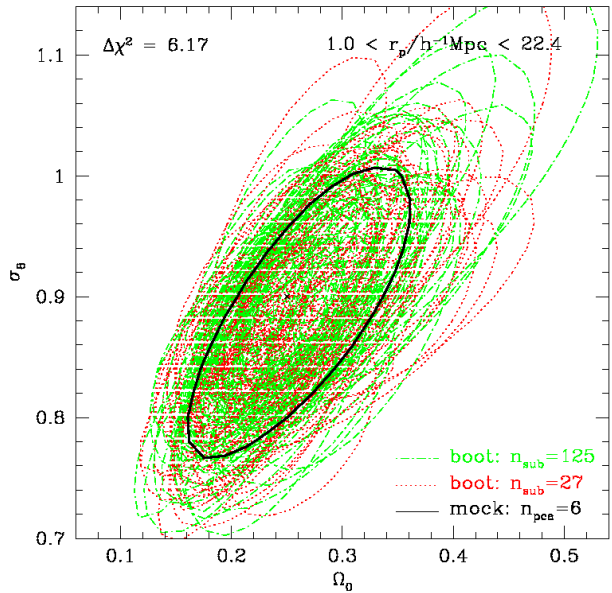}
\caption
{
  Same as Fig.~\ref{fig:om_sig_plane}, but for \nr=3~\nsub. The 
  ``true'' error estimate is given by the mock errors (thick black),
  while red (dotted) correspond to Boot-27 and green (dot-dashed) to
  Boot-125. This plot is a great contrast to the much messier
  Fig.~\ref{fig:om_sig_plane}. The fits are done over the projected
  scales indicated in the figure. See text for further comments.
}
\label{fig:om_sig_plane_nm}
\end{figure}

\begin{figure*}
\plottwo{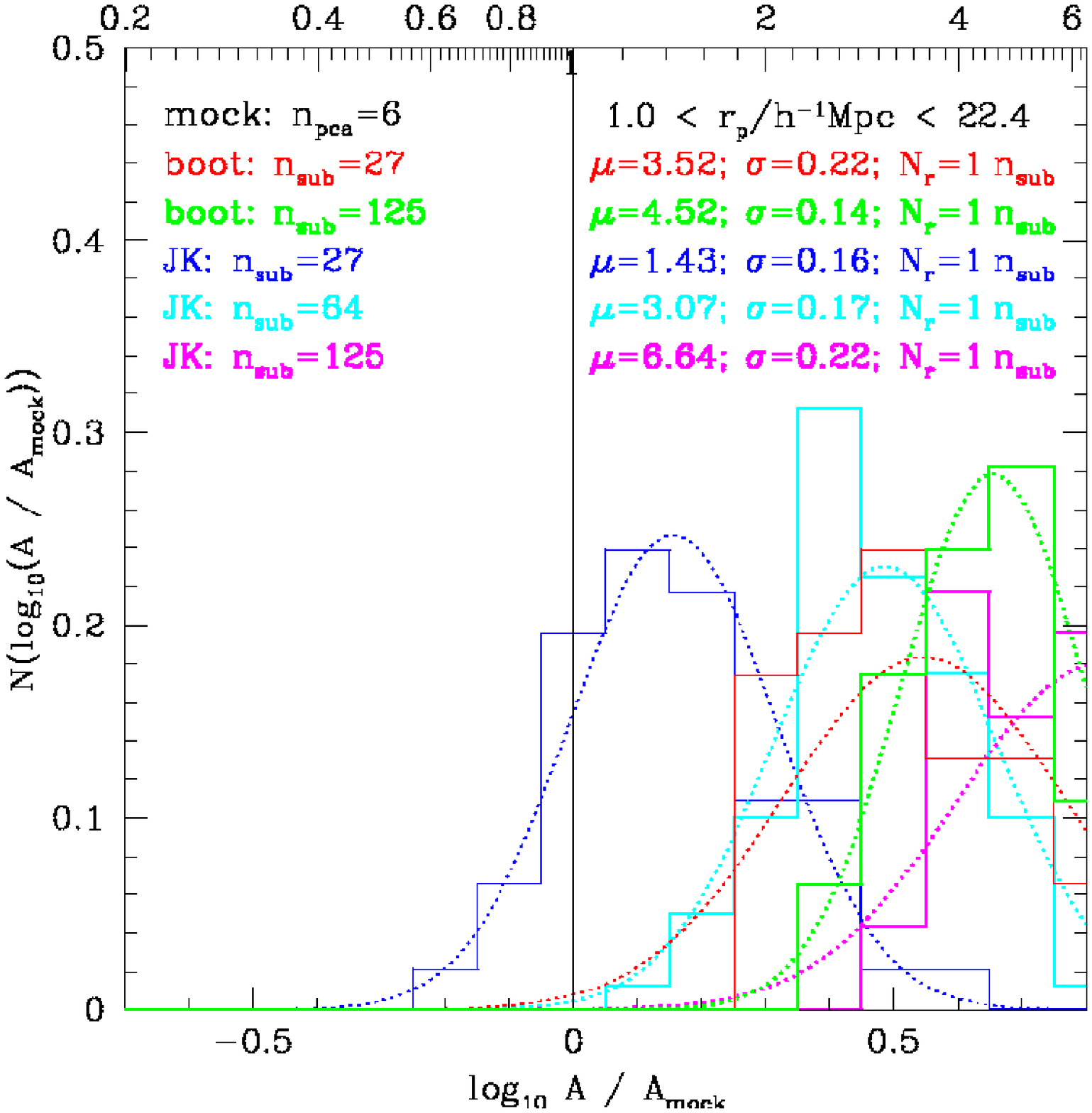}{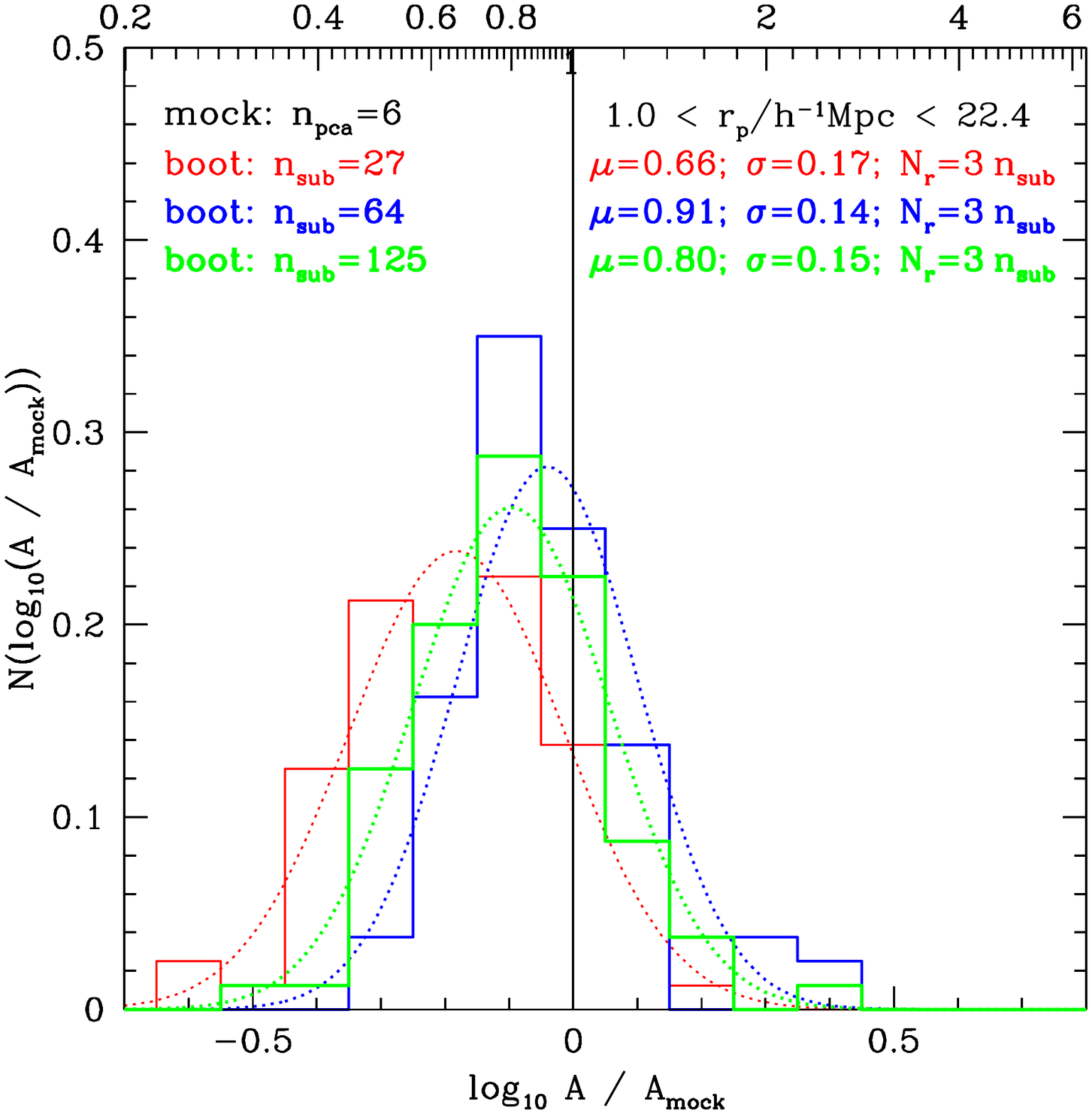}
\caption
{
The distribution of the figure of merit, similar 
to Fig.~\ref{fig:area_dist}, but for \nr=\nsub\ (left panel) and
\nr=3~\nsub (right panel), assuming a mean number density of 25~\%
\lstar\ (as opposed to a \lstar\ mean number density as in
Fig.~\ref{fig:area_dist}). The figure legend gives the exact meaning
of all the lines. The dotted lines are the corresponding Gaussian
distributions with similar variance and mean. Increasing the number of
sub-samples randomly drawn, with replacement, from the set of \nsub\
sub-samples radically changes the relative area distributions.
}
\label{fig:area_dist_nm}
\end{figure*}

So on adopting this essentially new prescription for the bootstrap error
analysis, i.e. allowing for bootstrap resamplings of typically three
times the number of sub-samples considered (\nr=3~\nsub), what is the 
impact on the other results presented in
Sections~\ref{sec:err_anal} to~\ref{sec:test_case}?
This question is best answered in Figs.~\ref{fig:om_sig_plane_nm}
and~\ref{fig:area_dist_nm}, which show the same information as
Figs.~\ref{fig:om_sig_plane} and~\ref{fig:area_dist}, but for
bootstrap estimates with \nr=3~\nsub.

Fig.~\ref{fig:om_sig_plane_nm} is a great contrast to the messy 
Fig.~\ref{fig:om_sig_plane}: this time internal error estimates seem to
return CI which are in much better agreement with the ``truth''. We
show, together with the mock CI in black, two sets of bootstrap
estimates, each applied to one hundred different data sets: Boot-27 (in
red) and Boot-125 (in green) with \nr=3~\nsub. We also note that the mock
contour is on average $\sim 50$~\% larger than in 
Fig.~\ref{fig:om_sig_plane}. This is due to the 25~\% lower number
density considered in this case, which is in good agreement with the
findings of Fig.~\ref{fig:boot_resampling}, for which we also found
a typical $\sim 50$~\% change in error w.r.t. the reference mean number
density. 

Fig.~\ref{fig:area_dist_nm} present the distributions of the relative
area statistic for a suite of internal error estimators, all estimated
over the range  1 to $\sim 22~\Mpc$, with N$_{\rm pca}=6$. In the left
panel, we show the distributions of the standard bootstrap estimates
with \nr=\nsub\ and those of the corresponding jackknife estimates:
Boot-27 in red, Boot-125 in green, Jack-27 in blue, Jack-64 in cyan and
Jack-125 in magenta. In the right panel, we show the distributions for
bootstrap with \nr=3~\nsub: Boot-27 in red, Boot-64 in blue and
Boot-125 in green. 

Fig.~\ref{fig:area_dist_nm} makes several points. 
First it becomes clear from the right panel that the number of
sub-samples drawn, with replacement, from a set of \nsub\ sub-samples
has to be typically three times larger than \nsub\ for bootstrap errors
to best reproduce the ``true'' mock inferred errors with minimal bias. 
Secondly, once that regime is reached, bootstrap errors are only 
very marginally dependent on the number of sub-samples
the data set is split into, i.e. once the criteria on the number of 
data points discussed in Hartlap \etal\ (2007) is satisfied. 
For our clustering statistics, this seems to be
true once \nsub\ is at least 5 times larger than N$_{\rm pca}$. 
Thirdly, the log-normal variance is limited to $\sigma \sim 0.15$. Hence
in the absence of bias, the area of the CI are known to within a factor
of 2 (with a 95~\% confidence level), explicitly meaning that an
estimated 2-$\sigma$ CI can be mistaken to be pessimistically a true
$\sim 1.3$-$\sigma$ or optimistically a true 3-$\sigma$. Finally the
observed systematic bias seen in Fig.~\ref{fig:area_dist_nm} is
actually less important than it looks, as 
the uncertainty on the mock error is larger for the number density
considered here. For $f=0.25$, we observe a $\sim 20$~\% scatter on
the mock error, which is more than twice as large compared to our
findings with $f=1.0$ (see \S\ref{sec:model_fit}).
 
Additionally, the left hand panel of Fig.~\ref{fig:area_dist_nm}, once
compared to Fig.~\ref{fig:area_dist}, shows that the dispersion on the
internal error estimates seem to be rather insensitive to the actual
number density of the samples analysed. Decreasing the sampling rate
down to 25~\% of the original (as in Fig.~\ref{fig:area_dist_nm}) makes
barely any difference to the quoted dispersions, as long as the
distributions of relative areas are still well described by
Gaussians. On the other hand, the bias seems to be a much stronger
function of the assumed mean number density, with lower number 
densities tending to systematically overestimate the errors by
quite a significant amount. This is explicitly shown by the large mean
values of each relative area distribution shown in the left panel,
ranging from $\sim 1.3$ to as large as $\sim 6.6$ in the case of
Jack-125. The bias increases by 1.5 to 2.5 times, depending on the
estimator, when the mean density becomes 4 times smaller. When the bias
becomes so large and the changes so unpredictable, it is no longer
clear whether there is any positive side to be seen with these
traditional internal estimators.

\section{Summary and Conclusions}
\label{sec:conclusion}

In this paper we have carried 
out an extensive comparison of the relative performance 
of internal and external error estimators for two-point clustering 
statistics. We devise a set of numerical experiments, extracting 
independent data sets from N-body simulations. The data sets are 
chosen to have a volume comparable to that of the typical \lstar\
volume limited samples constructed from the SDSS. The benchmark for
this exercise is the error estimate obtained from the scatter over our
independent data sets, which we refer to as ``mock'' errors or
``truth''. This is then compared to internal estimates made using the
jackknife and bootstrap techniques (e.g. Tukey 1958; Efron 1979). We
revisit the assumptions and the free parameters behind these techniques
to see if we can lay down a recipe that would reproduce the more
expensive (and not even always possible) external errors. We summarize
below our findings, starting with uncorrelated statistics, followed by
covariance matrix based results and ending with the conclusions drawn
from a case study, aimed at fitting two cosmological parameters to our
clustering results.

Perhaps surprisingly (as they are so widely used in the literature),
we find that both internal estimators of the errors have worrying
failings. Neither was able to faithfully reproduce the relative
variance of the external estimate, over a limited range of scales (0.5
to 25$~\Mpc$). At face value, the standard bootstrap variance is nearly
50~\% larger on all scales, where standard refers to the case for which
the number of sub-samples selected at random, with replacement, equals
the number of sub-samples the data set is divided into
(i.e. \nr=\nsub). In \S\ref{sec:boot_resampling}, we solve this problem
of the overestimated variance simply by increasing \nr\ by a factor of
three, i.e. by setting \nr=3~\nsub. The variance measured with the
jackknife method is fairly accurate on large scales (greater than
$10~\Mpc$), but the variance on small scales (less than 2-3~$\Mpc$) is
clearly overestimated by a significant amount and in a method
dependent way: the bias depends strongly on the number of 
sub-volumes the data set is split into, as explicitly shown in
Fig.~\ref{fig:variance}. Another major result from
Section~\ref{sec:error_dist} is that all error estimators considered
present Gaussianly distributed errors on the smallest projected
separations. However, slowly but surely, the error distributions become
non-Gaussian already on 6~(15)~$\Mpc$ and larger for the jackknife
(bootstrap) method, while they all remain Gaussian for the mocks over
the full range of scales considered.

The details of the recovered covariance matrices, or more precisely
the recovered principal component decomposition of the covariance
matrices for our 100 data sets, show equally worrying features in
Sections~\ref{sec:eigen_val} and~\ref{sec:eigen_vec}.
Generally speaking bootstrap inferred (normalized) eigenvalues
and eigenvectors are in good agreement with mock inferred ones,
while jackknife inferred (normalized) eigenvalues and eigenvectors
present distinctive features which are not present in the mocks and, 
furthermore, are also dependent on the number of sub-volumes the data 
set is split into. These features are particularly sensitive to the smallest
scales used: as for the variance estimates, the jackknife method
becomes increasingly discrepant w.r.t. the mocks when scales smaller
than 2-3~$\Mpc$ are used. However, the direct influence of those
subtle differences in a such technical analysis is hard to grasp and
hence a better understanding in reached on an example scenario.

In Section~\ref{sec:test_case} we present a case study in which the
different error estimation methods are used to extract constraints on
cosmological parameters from measurements of the projected correlation
function, using 100 independent data sets. The discrepancy in the
relative variance returned by the 
different methods, as described above, propagates through this analysis
and gives different constraints on the fitted parameters.  
We quantify these differences in terms of the area of the confidence
interval which would correspond to 2-$\sigma$ for the case of a two
parameter fit for Gaussian distributed measurements. With 100
independent data sets, we find that the internal estimators return, on
average, error ellipses with larger area than that found in the mocks,
resulting indirectly into a redefinition of the confidence limit of the
measurement. This is particularly true for standard bootstrap
estimates, for which the number of sub-samples selected at random with
replacement equals the number of sub-samples the data set is divided
into (i.e. \nr=\nsub). However, we show in \S\ref{sec:boot_resampling}
that increasing the number of sub-samples drawn at random by a factor
of three (i.e. \nr=3~\nsub) solves most problems: in that case, the
confidence intervals are only marginally different to the mock ones. 
In the case of jackknife errors, the area of the error ellipse is to
some extent sensitive to the number of sub-volumes the data set is
split into. For all error estimators, we find, as expected, 
that the error ellipse area is sensitive to the number of principal
components used in the analysis.

The diagnosis for the internal estimators is therefore mixed. The
jackknife method has problems recovering the scale dependence of errors
and the results are sensitive to the number of sub-samples the data set
is split into. There is little scope to fix these problems given the
definition of the method; the only thing we can vary is the number of
sub-samples into which the data set is split. We did not find one
choice for the number of sub-samples which could cure all of the
ailments of this method. The prognosis for the bootstrap is on the
other hand more encouraging. The problem of overestimating the variance 
can be traced to the effective volume of the data set used when the number 
of sub-volumes chosen at random with replacement is equal to the number 
of sub-volumes the data set is divided into. By oversampling the sub-volumes, 
this problem can be fixed, with the effective volume used tending to the 
original data set volume. Better still, for our application at least, 
there appears to be an optimal factor, three times, to oversample, 
with higher rates producing too little variance.

Unfortunately there seems to be no hard and fast rules for the 
best way to set about a principal component analysis of the covariance 
matrix of clustering measurements. The value of the principal component 
analysis is that it helps break down the information contained in a 
clustering measurement. The measurement will be expressed in a
restricted number of bins. The choice of the number of bins is
arbitrary. Using more bins does not necessarily imply that there will
be more information in the correlation function. The PCA breaks the
covariance matrix down into eigenvectors. These are ranked in terms of
how much information or variance they contain. The variance drops
quickly with the order of the eigenvector for the examples we
considered, indicating that most of the information in the clustering
measurements can be broken down into a few terms. 
The best advice we can give here would be to compare the results obtained 
using different numbers of eigenvectors and choose a value where the results 
and conclusions do not change significantly.   

The analysis presented in this paper is applicable to any galaxy or
cluster survey with three dimensional information. Some of the issues
discussed relating to redshift space distortions 
are particular to local surveys in which the distant observer
approximation does not hold.
A new set of experiments would be required to extend our results to the 
calculation of errors for photometric surveys, which look at angular
clustering, or to multi-band photometric surveys, which will use
photometric redshifts to look at clustering in redshift slices. The
projection involved in these catalogues changes the underlying
statistics, making them look more Gaussian perhaps. This is the most
likely explanation why Cabr\'e \etal\ (2007) find such a good agreement
between jackknife and mock errors for angular clustering statistics.

\section*{Acknowledgements} 
 
PN wishes to acknowledge numerous stimulating discussions with
Cristiano Porciani, Martin White, Idit Zehavi as well as many other
participants at the 2006 and 2007 Aspen summer workshops, and the kind
use of many computers at the IfA and the ICC. 
PN is supported by a PPARC/STFC PDRA fellowship Fellowship. 
EG acknowledge support from Spanish Ministerio de Ciencia y Tecnologia
(MEC), project AYA2006-06341 and research project 2005SGR00728 from
Generalitat de Catalunya. 
CMB is supported by a Royal Society University Research Fellowship. 
DC acknowledges the financial support from NSF grant AST00-71048.
This work was supported by the EC's ALFA-II programme via its funding of
the Latin American European Network for Astrophysics and Cosmology.
The {\tt L-BASICC} simulations in this paper were carried out by Raul Angulo
using the Virgo Supercomputing Consortium computers based at the
Institute for Computational Cosmology at Durham University.

\end{document}